\documentclass[twocolumn,english,prl]{revtex4-1}
\usepackage{xcolor}
\usepackage{graphicx}
\usepackage[T1]{fontenc}
\usepackage{float}
\usepackage{textcomp}
\usepackage{amsmath}
\usepackage{amssymb}
\usepackage{graphicx}
\usepackage{esint}
\usepackage{natbib}
\usepackage{color}

\begin{document}

\title{Topologically protected wave packets and quantum rings in silicene}

%% Notice placement of commas and superscripts and use of &
%% in the author list

\author{Bart\l{}omiej Szafran, Bart\l{}omiej Rzeszotarski, and Alina Mre\'nca-Kolasi\'nska}

\affiliation{AGH University of Science and Technology, Faculty of Physics and
Applied Computer Science,\\
 al. Mickiewicza 30, 30-059 Krak\'ow, Poland}

\begin{abstract}
We study chiral wave packets moving along the
zero-line of a symmetry breaking potential of vertical electric field in buckled silicene
using an atomistic tight-binding approach with initial conditions set by an analytical solution of the Dirac equation.
We demonstrate that the wave packet moves with a constant untrembling velocity and with a preserved shape along the zero line.
Backscattering by the edge of the crystal is observed that appears with 
the transition of the packet from $K$ to $K'$ valley or vice versa. 
We propose a potential profile that splits the wave packet and next
produces interference of the split parts  that acts as a quantum ring. 
The transition time exhibits Aharonov-Bohm oscillations in the external magnetic field that are translated to conductance oscillations when
the intervalley scattering is present within the ring.
{ We study wave packet dynamics as function of the width of the packet to the limit of plane waves. In the stationary
limit the conductance oscillation period is doubled and  the scattering density oscillates between the left and right arms of the ring as function of the magnetic field. 
We demonstrate that this effect is also found in a quantum ring defined by the zero lines in bilayer graphene.}
\end{abstract}

\maketitle

\section{Introduction}

In monolayer honeycomb materials, including  graphene \cite{gra}, Xenes  \cite{Molle17} (silicene \cite{Aufray10,Liu11}, germanene \cite{cha,Liu11}, stanene \cite{sta}),
transition metal dichalcogenides \cite{tmdc}, or bismuthene \cite{Reis17}, the Fermi level appears in two non-equivalent valleys of the Brillouin zone. 
In graphene ribbons \cite{nrbr} with zigzag edges there is a strict correspondence between
the valley state and the direction of the current flow \cite{zz1,zz2,zz3}. 
The valley protects the chiral \cite{chiral} electron transport from backscattering by long
range potential disorder, which led to the
valley electronics \cite{rycerz,ryre} or application of the valley degree of freedom to information processing \cite{ryre,vq1,vq2}. 

In bilayer \cite{morpugo} and staggered monolayer graphene \cite{niu}  a topological confinement inside the sample, far from the edges, was  found \cite{morpugo,niu} along a zero line of inversion-symmetry-breaking potential.  
For bilayer graphene \cite{bgr} this potential is introduced by an electric field perpendicular to the layers \cite{morpugo}. The bias opens the energy gap in the band structure \cite{bgr,wry}.
 For an inhomogeneous electric field that is inverted at the zero line
in space,  topologically protected chiral currents have been found \cite{morpugo}
 confined in space to a region of the electric field flip. The flip line provides a one-dimensional confinement or quantum-wire-type channels \cite{macdo,peet,muktu,muktu2}.
The one-dimensional confinement of currents is also found in the quantum Hall conditions
at the n-p junctions \cite{np} induced by electric fields. In contrast to the currents localized at the n-p interface \cite{rick,tay,snejk}, the confinement in zero-line channels does not require external magnetic field \cite{morpugo}. { Note, that formation of an energy gap was also found in epitaxial monolayer graphene 
due to the sublattice symmetry breaking by the substrate \cite{sic}. However, the gap of this 
origin cannot be easily inverted for the topological confinement.}

A perpendicular electric field opens the energy gap for the buckled Xenes monolayers \cite{Molle17,ni,Drummond12}, and an inversion of the field induces topological confinement similarly as 
in bilayer graphene \cite{Ezawa12a}. In contrast to bilayer graphene, 
(i) in Xenes the chiral energy level 
that passes from the valence to the conduction band 
is a linear function of $k_y$. 
In Xenes a (ii) single topological state per current direction is
present instead of two as in bilayer graphene. 
As a consequence of (i) and (ii) the wave packet formed at 
zero line in a Xenes should be stable against excitations and should travel
with a constant shape due to independence of the velocity on the wave vector.

In this paper we study the dynamics of the chiral wave packets along the zero lines
of the electric field in silicene \cite{Molle17,rivi,chow,Ezawa,s1,s2,s3}, which is the most advanced Xenes material, with 
a successful application for the field effect transistor \cite{Tao15}.
The chiral wave packets move with the Fermi velocity and the motion of the topological packets is free from
Zitterbewegung \cite{ci2,citer,frolo,naturd} which is characteristic to the solution of the Dirac equation
and was found also for Weyl fermions in graphene \cite{citer,rome,frolo}.    
Moreover, the topological electron packets move with a constant shape, similarly to solitons,
that in other conditions require interaction
with the environment --  see the electron solitons self-focused
with interaction to the metal gates in heterostructures \cite{bed} or the
 Trojan wave packets formed by carefully prepared electromagnetic field \cite{tro1,tro2}.

We show that the chiral electron packets can be transferred from one valley to the other by backscattering from the edge of the crystal and we find that the packet appears in the opposite valley in a restored shape.
Potential profiles that act as beam splitters and quantum rings \cite{qr}
 are proposed. 
An interference of the split parts of the wave packet can be controlled by external
magnetic field threading the quantum ring. The ring stores the wave packet for a time that is 
a periodic function with the period the  flux quantum. The Aharonov-Bohm \cite{ab} conductance oscillations are also demonstrated  for the system with an intervalley scattering present. 
The topological rings are a new form of quantum rings for Dirac electrons, with respect
to previously considered systems with structural confinement \cite{structure0,structure,structure2,structure3,structure4},
mass confinement \cite{massconf,massconf2,massconf3} or the 
confinement at circular n-p junctions  %\cite{rick,tay,snejk} 
in the quantum Hall conditions \cite{amk}.

{
In the long wave packet limit, i.e. when the width of the packet is comparable with the 
diameter of the ring, the interference within the ring is translated into an imbalance of the 
scattering density in the left and right arms of the ring that becomes a periodic function 
with doubled Aharonov-Bohm period. The result is also found in the stationary transport
for silicene and bilayer graphene. }

This paper is organized as follows.
In Section II we provide an analytic solution to the Dirac equation for the 
topologically protected currents at the inversion
of the electric field (II.A). In section II.B the solution of the continuum approximation (II.A)
is translated to atomistic tight-binding description which naturally accounts 
for the intervalley scattering, and the time stepping procedure is explained.
In the Section  III we  test the stability of the wave packet
against the intervalley transition (III.A) and  describe the Aharonov-Bohm \cite{ab} oscillations
of the electron storage time by the ring (III.B). Section III.C describes the conductance oscillations for intervalley scattering within the ring.
{ The long wave packet limit is disussed in III.D, and the limit
 is confronted with the standard  calculation based on the stationary electron scattering in Section III.E. Section III.F shows the results of the Landauer approach for 
quantum rings defined by the zero lines in bilayer graphene. 
 Section IV contains the summary.}

\begin{figure}
\begin{tabular}{llll}
(a) &\includegraphics[width=0.5 \columnwidth]{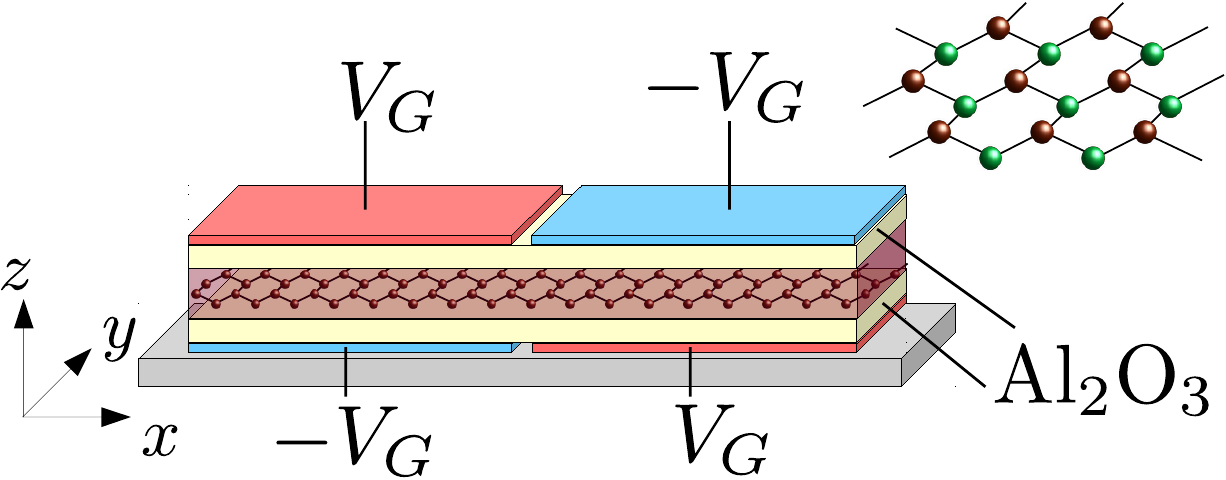} &
(b) &\includegraphics[width=0.3\columnwidth]{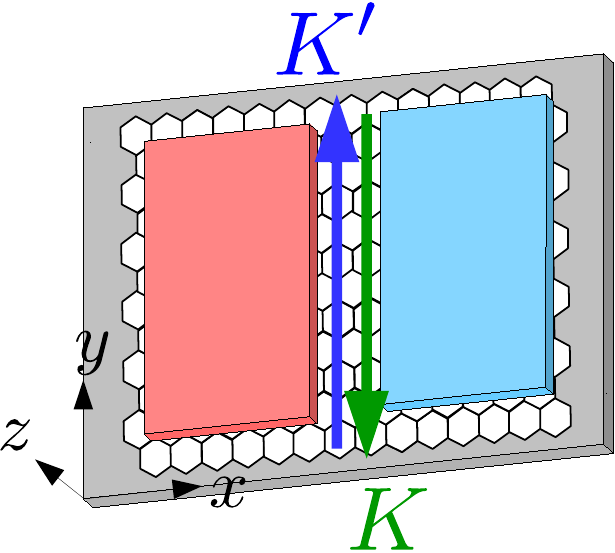} \\
(c) &\includegraphics[height=0.6\columnwidth]{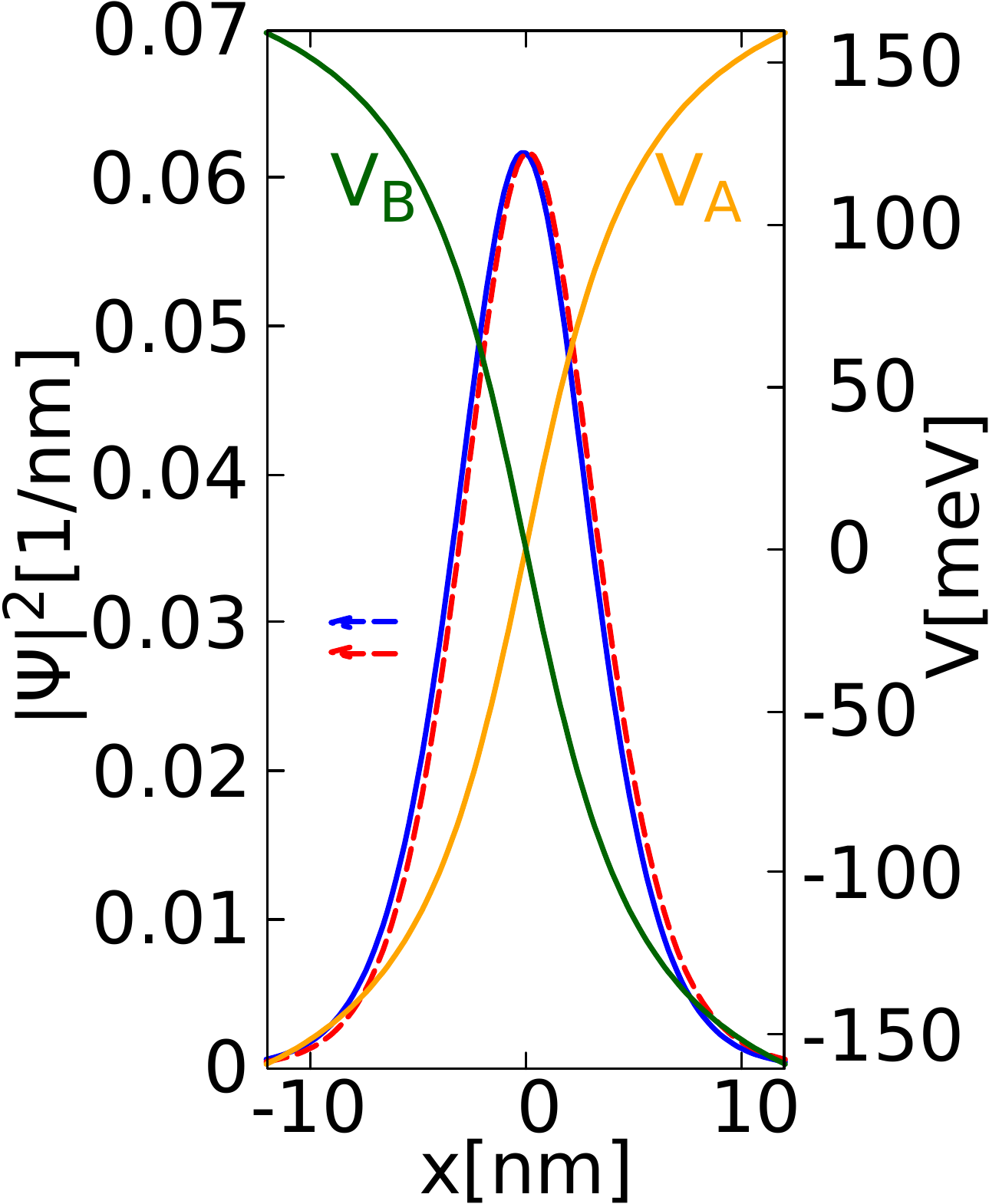} &
(d) &\includegraphics[height=0.6\columnwidth]{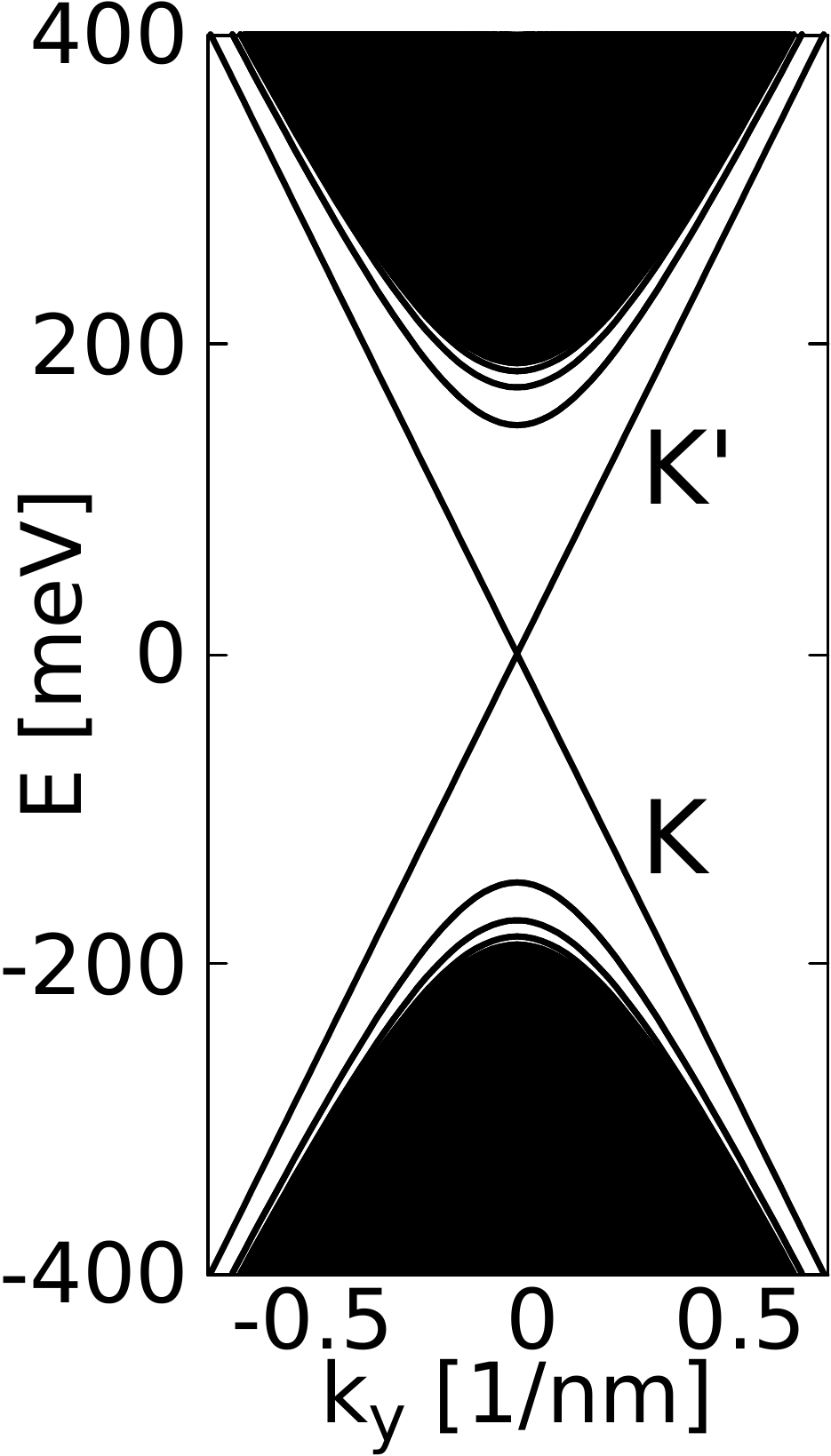}% \put(-18,50){$K$}\put(-21,100){$K'$} 
\end{tabular}
\caption{(a) Schematic side view of a silicene monolayer embedded in a dielectric sandwiched
between top and bottom gates with the A sublattice on top (red dots in the inset)  closer to a positively $V_G>0$ biased top gate (red color) gated inducing a negative potential energy  for $x<0$. (b) Top view of the system. The $K'$ ($K$) valley  current flows leaving the negative (positive) potential energy  on the A sublattice on the left hand side. The flake has a zigzag (armchair) termination at the edges with constant $y$ ($x$). 
(c)  The potential profile on the A and B sublattices (right axis) and the wave function on the 
A sublattice as calculated in the continuum approximation [Eq. (6)]. The blue solid (red dashed) lines
show the wave functions with the positive (negative) product of the valley index $\eta$ 
and the spin $\sigma_z$ eigenvalue for $\lambda=4$ nm and $V_g=0.2$ eV. (d) The dispersion relation 
calculated numerically. 
 \label{sze}}
\end{figure}

\begin{figure}
\includegraphics[width=.6\columnwidth]{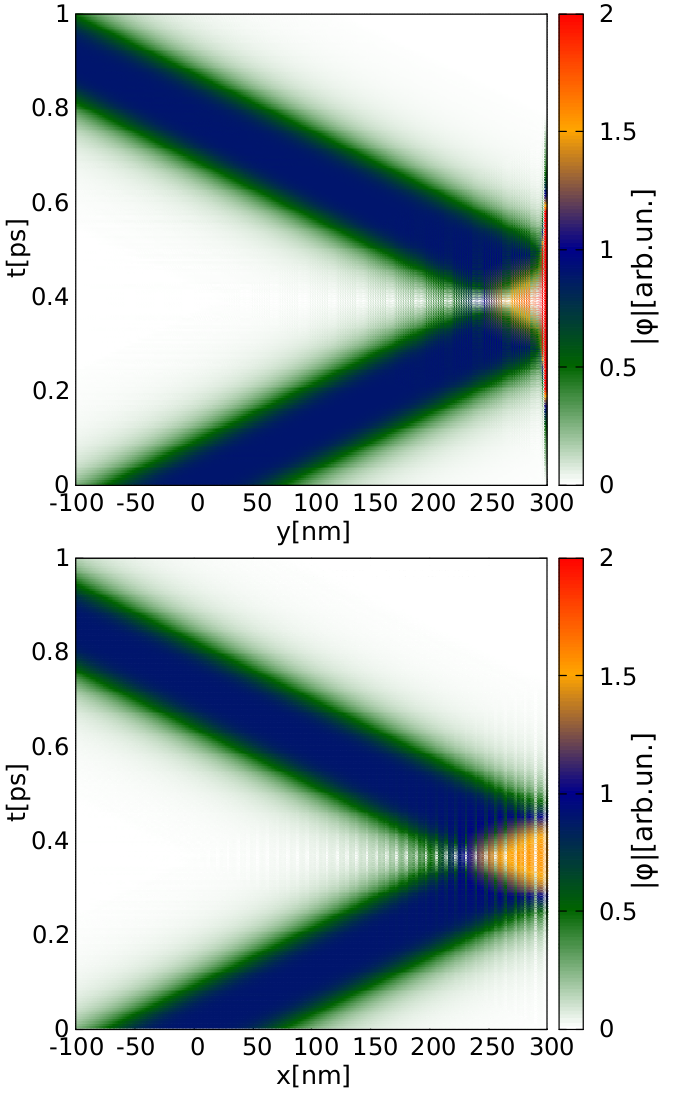}
\put(-50,225){(a)} \put(-50,100){(b)}
\caption{Time evolution of the absolute value of the electron packet ($|\phi|$) backscattered by the zigzag (a) and armchair (b) 
edge of the flake. The square flake has a zigzag edge  at $y=300$ nm (a) and an armchair edge at $x=300$ nm (b). The plots show the absolute value of the wave function
at  the inversion of the energy band that is introduced at $x=0$ (a) and at $y=0$ (b).
For the zigzag edge the wave packet is temporarily localized at the edge.
The vertical fringes observed
at the backscattering  result from the superposition of the incoming and outgoing waves from the opposite valleys  $K$ and $K'$.
}
 \label{aring}
\end{figure}

\section{Theory}

\subsection{Chiral currents confined by the energy gap inversion}
The chiral currents at the band inversion in silicene similar to the ones in bilayer graphene  \cite{morpugo} were found in Ref. \cite{Ezawa12a}.
For buckled monolayers as well as for bilayer systems the energy gap can be tailored in space
using multiple split gates with inverted polarization. The idea of the local manipulation of the energy gap by dual split gates  was proposed for topological confinement \cite{morpugo} 
and pseudospin electronics in bilayer systems \cite{re2a,re2b,re2c}. 

Here, we consider a buckled silicene monolayer sandwiched in between top and bottom gates  [Fig. \ref{sze}(a,b)].
  The gates are split, so that the electric field changes orientation along the $y$ axis (i.e. for $x=0$).
We model the potential at the A sublattice using an arctangent function
\begin{equation}
 V_A(x)=\frac{2V_g \arctan(x/\lambda)}{\pi}. \label{pote}
\end{equation}
%$|V_A(x)|$ attains $0.9V_g$ for $|x|=6.31\lambda$.
We assume that the silicene is embedded symmetrically between the gates, so that 
on the B sublattice we have  $V_B(x)=-V_A(x)$  [Fig.\ref{sze}(c)]. 
The potential bias between the sublattices opens an energy gap in the band structure \cite{ni,Drummond12}.
For potential of Eq. (\ref{pote}) the energy gap is inverted at $x=0$ by the flip of the electric field orientation.% [Fig. \ref{sze}(c)].

For the wave function components defined on sublattices 
$\psi=\left( \psi_A , \psi_B\right)^T$  the low-energy approximation to the atomistic tight-binding Hamiltonian  reads \cite{Ezawa12a}
\begin{equation} H_\eta = \hbar v_F \left(k_x \tau_x -\eta k_y \tau_y \right)+V({\bf r})\tau_z
%+\frac{g\mu_B B}{2}\sigma_z
 -\eta \tau_z \sigma_z  t_{SO},\label{hamu} \end{equation}
where $\eta$ is the valley index ($\eta=1$ for the $K$ valley and -1  for the $K'$ valley), 
 $\tau_x$, $\tau_y$ and $\tau_z$ are the Pauli
matrices in the sublattice space, ${\bf k}=-i\nabla$, the Fermi velocity $v_F$ is determined
by  the  nearest distance between Si atoms
$d=2.25$ \AA\;  and the tight-binding hopping parameter $t=1.6$ eV \cite{Liu11,Ezawa}
 $v_F=3dt/2\hbar$.  In Eq. (\ref{hamu}) $t_{SO}=3.9$ meV is the intrinsic spin-orbit coupling constant \cite{Liu11,Ezawa}. 
%In the Hamiltonian (\ref{hamu}) we introduced an assumption
%that the potential $V({\bf r)}$ is opposite at both the sublattices, which implies that the silicene
%plane is inserted symmetrically in a system of top and bottom electrodes. 
The intrinsic spin-orbit coupling is diagonal in the basis of eigenstates of the $z$ component of the spin and $\sigma_z=\pm 1$ is treated as a quantum number. 

%In the following we will consider a perpendicular magnetic field $(0,0,B)$ that will be introduced by the vector potential of the form ${\bf A}=(0,Bx,0)$. 
With potential given by Eq. (1) the Hamiltonian commutes with the $\frac{\partial}{\partial y}$ operator. The common eigenfunctions of the energy and $y$ momentum component can be put in form
$\psi^{k_y}(x,y)=\exp(ik_yy)\left(\begin{matrix} \psi_A^{k_y}(x) & \psi_B^{k_y}(x)\end{matrix}\right)^T$. %For fixed $k_y$,
%For the system without the external magnetic field ${\bf A}=0$ 
%We provide analytical solutions to the Schroedinger equation for the states localized at the flip of the electric field.
The Hamiltonian eigenfunctions
% components 
fulfill the % following
 system of equations
\begin{eqnarray} \frac{(V_A(x)-E-\eta\sigma_z t_{SO})\Psi_A(x)}{i\hbar v_F}&=&(\Psi_B'-\eta k_y \Psi_B), \\
\frac{(-V_A(x)-E+\eta\sigma_z t_{SO})\Psi_B(x)}{i\hbar v_F}&=&(\Psi_A'+\eta k_y \Psi_A),
\end{eqnarray}
where the prime stands for $x$ derivative. % $\frac{\partial}{\partial x}$.
 We plug in Eqs. (3,4) a relation $\Psi_B=i\Psi_A$, i.e.  a guess based on a numerical solution that allows us to derive an analytical solution for states localized near the zero-line  area. 
A sum of the resulting equations relates the potential and the wave function  \begin{equation} V_A(x)-\eta \sigma_z t_{SO}=-\hbar v_F \frac{\partial}{\partial x}\ln \Psi_A(x).\end{equation}
For the specific form of potential given by Eq. (1) the (unnormalized) wave function is found by a standard integration technique
\begin{equation}\Psi_A=(\lambda^2+x^2)^{\frac{\lambda V_g}{\pi\hbar v_F}}e^{(-\frac{2V_g x}{\pi \hbar v_F}\arctan\frac{x}{\lambda}+\frac{\eta\sigma_z t_{SO} x}{\hbar v_F})}.\end{equation}
The term $x\arctan(x/\lambda)$ in the exponent for low $x$ introduces a Gaussian-like dependence
to the wave function and keeps it localized near the band inversion area [Fig. \ref{sze}(c)]. 
The average value of $\langle x ^2\rangle^{1/2}$ for $\lambda= 1$, 2, 4, 12, and 20 nm is 
2.45, 2.87, 3.55, 5.45 and 6.83 nm, respectively.
%The localization at the edge increases with $V_g$ and decreases with $\lambda$. 
In presence of the spin-orbit coupling the wave function is not ideally symmetric
with respect to the center of the gap inversion line: the term including 
the spin-orbit coupling shifts the wave function at left (right) of the inversion line for
negative (positive) product of valley index $\eta$ and the $\sigma_z$ eigenvalue.
For the applied parameters these shifts are not very strong -- see the blue and red lines in Fig. \ref{sze}(c). 
%The localized wave function of Eq. (6) is exactly reproduced by the finite difference calculations. 
 In the calculations below we set $\sigma_z=1$.

\begin{figure}
\begin{tabular}{ll}
(a) &\includegraphics[height=0.4\columnwidth]{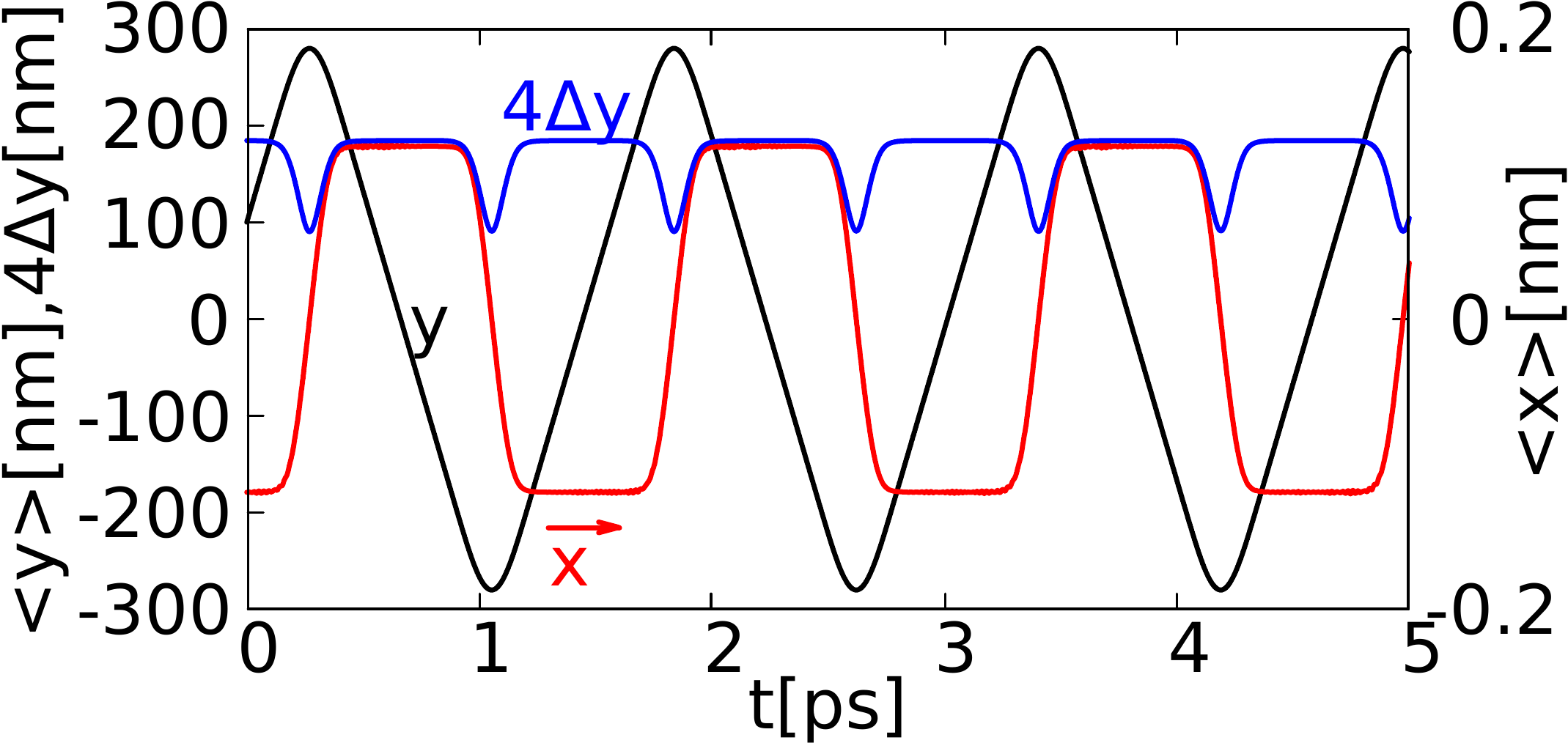}  \\
(b) &\includegraphics[height=0.4\columnwidth]{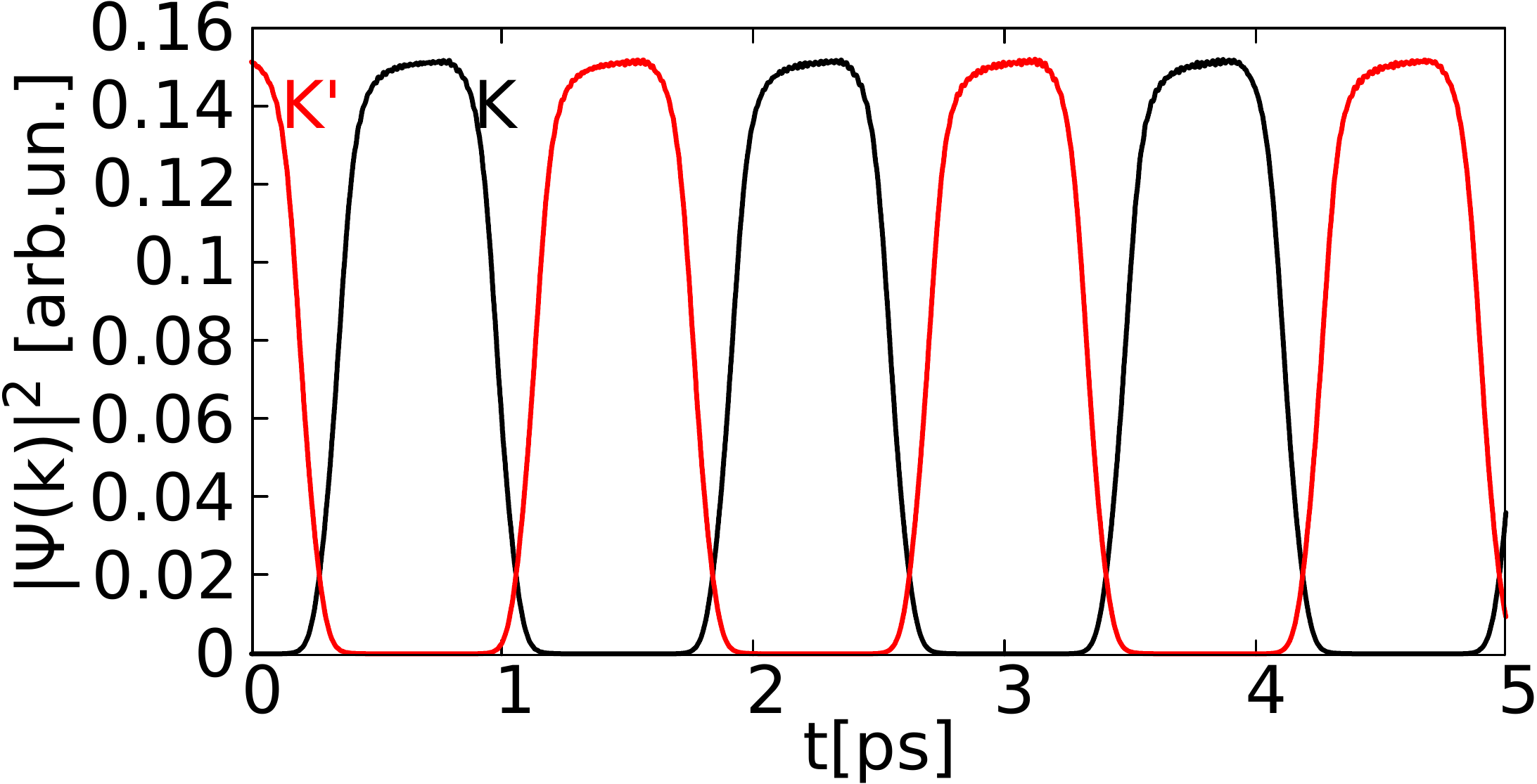}% \put(-18,50){$K$}\put(-21,100){$K'$} 
\end{tabular}
\caption{(a) The average position in the $y$ (black line, left axis) direction for the wave packet
scattered by the armchair edge of the flake. The spread of the wave packet $\sqrt{\langle (y-\langle y \rangle )^2\rangle}$ multiplied by 4 (blue line, left axis), and the average position at the $x$ axis
(right axis, red line). The variation of $\langle x\rangle $ are due to the spin-orbit coupling and the valley transitions that follow the backscattering by the edge [see. Eq.(6)]. The electron spin is set at $\sigma_z=+1$.
The parameters of the system are the same as in Fig. 2(a), only the packet is started with the initial
average position at $y_0=100$ nm.
(b) The Fourier transform of the packet calculated for the $K'$ (red) and $K$ valley (black)
} \label{wyna}
\end{figure}

The energy of the states localized at the gap inversion can be calculated by adding Eqs. (3) and (4) still with the relation $\Psi_B=i\Psi_A$, which gives  \begin{equation} E=-\eta \hbar k_y v_F. \end{equation}
The entire dispersion relation calculated numerically with a finite difference approach \cite{sweep} is given in Fig. 1(d).
The linear band energy is independent of the $V_g$ or $\lambda$ which only affect
the transverse  wave function localization at the energy gap inversion line. 
 Above the energy gap a continuous spectrum is found with a parabolic threshold as a function of $k_y$. 
Near the zero energy only the localized reflectionless currents flow, and for $V_g=200$ meV,  the gap is wide enough to make the currents stable at room temperature. For the Fermi wave vector $k_F\simeq 0.25$ nm$^{-1}$ ($E_F\simeq 100$  meV) the continuum appears still 100 meV above the linear band.

The sign of the electron velocity within the linear band $v(k_y)=\frac{1}{\hbar}dE/dk_y=-\eta v_F$ depends
on the valley index.
 Hence, the transport at the inversion line is chiral, i.e. the electron states of the valley $K'$ ($K$) go up (down) along the electric field flip line [Fig. \ref{sze}(b)].
Generally, in the states localized along the field flip the current in  the $K'$ ($K$) valley flows with the negative (positive) potential at the A sublattice at the left-hand side.
%The currents flowing along the electric-field flip 
%are protected by the valley degree of freedom and the backcattering is only possible with the intervalley transition which can only be induced by abrupt potential changes and not a smooth potential variation.

\subsection{Chiral solitons in the atomistic tight-binding}
\subsubsection{Atomistic Hamiltonian}

Since the electron velocity in the linear chiral band is independent of $k_y$ the wave packet localized  at the flip of the electric field should move with an unchanged shape along the $x=0$ line.   The description of the electron wave packets stabilized by the valley degree of freedom calls for an approach that takes into account the intervalley scattering. A natural choice is the atomistic tight-binding approach.
The positions of the ions of the A sublattice 
${\bf r}_{m}^A=m_1 {\bf a}_1+m_2 {\bf a}_2$  are generated 
with the crystal lattice vectors
${\bf a}_1=a \left(\frac{1}{2},\frac{\sqrt{3}}{2},0\right)$
and ${\bf a}_2=a \left(1,0,0\right)$, where $a=3.89$ \AA\; is the silicene lattice constant, and $m_1$, $m_2$ are integers. 
The B sublattice ions are generated by ${\bf r}_{m}^B={\bf r}_{m}^A+(0,d,\delta)$,
with the vertical shift of the sublattices $\delta=0.46$ \AA.
The coordinates of the center of the valleys in the reciprocal space 
are ${\bf K}_\eta=\left(\frac{4\pi\eta}{3a},0\right)$ \cite{rivi}. The valleys for $\eta=1$ (-1) are
referred to as $K$ ($K'$).

We use
 the Hamiltonian \cite{Liu,Ezawa,chow} 
\begin{eqnarray}
H_{TB}&=&-t\sum_{\langle m,l\rangle } p_{ml} c_{m}^\dagger c_{l} 
+it_{SO} \sigma_z\sum_{\langle \langle m,l\rangle \rangle  } p_{ml} \nu_{ml} c^\dagger_{m} c_{l} \nonumber 
\\ && +\sum_{m} V({\bf r}_m) c^\dagger_{m}c_{m}+\frac{g\mu_B B}{2}\sigma_z,  \label{hb0}
\end{eqnarray}
where $\langle m,l\rangle $ stands for the nearest neighbor ions,
$\langle\langle m,l\rangle\rangle $ for the  next-nearest-neighbor ions.
For the potential $V({\bf r}_m)$ we take $V_A({\bf r}_m)$  or $V_B({\bf r}_m)$.
The sign  $\nu_{ml}=\pm 1$ is  plus (minus)  for   the next nearest neighbor hopping path
 via the common neighbor ion that turns
 counterclockwise (clockwise). In Eq. (8) $p_{ml}$ is the Peierls phase 
 $p_{ml}=e^{i\frac{e}{\hbar}\int_{\vec{r_m}}^{\vec{r_l}}\vec {A}\cdot \vec {dl}}$,
where $\vec{A}=(0,B x,0)$ is the vector potential, and $B$ is the value of the magnetic 
field that is oriented perpendicular to the silicene plane. The last term in Eq. (8) is the spin Zeeman term with the Land\'e factor $g=2$, and Bohr magneton $\mu_B$.

\subsubsection{Initial condition and the time-stepping}

In the calculations to follow for the initial condition 
we use  the solution of the continuum Hamiltonian [Eq. (2)] and localize 
the packet along the band inversion using an envelope of form $\frac{1}{1+\frac{(y-y_0)^4}{D^4}}$,
with $D=80$ nm and $y_0$ that sets the center of the packet. We set the valley momentum with a plane wave and the $K$ or $K'$ coordinates.
Accordingly, for the atoms of the A sublattice we set as the initial condition
\begin{equation}
 \psi({\bf r}^A_m,t=0)=\exp\left(i ({\bf K_\eta}+{\bf k}_y) \cdot {\bf r}_m^A\right)\chi(y_m^A), \end{equation}
where ${\bf k}_y=(0,k_y,0)$ is the wave vector of the packet calculated with respect to the valley center, and $\Psi_A$ is given by Eq. (6).
For the atoms on the $B$ sublattice we take
\begin{equation}
 \psi({\bf r}^B_m,t=0)=i\eta \exp\left(i ({\bf K_\eta}+{\bf k}_y) \cdot {\bf r}_m^B\right)\chi(y_m^B), \end{equation}
where $\chi(y)=\frac{\Psi_A(y)}{1+\frac{(y-y_0)^4}{D^4}}$.
We set $k_y=0$ unless stated otherwise.

We solve the Schr\"odinger equation on the atomic lattice
$i\hbar \frac{\partial \psi} {\partial dt}=H\psi$, using the time step of $dt=10$ atomic units
or $dt=2.418\times 10^{-4}$ ps.
The wave function at the first step is calculated with the
implicit Crank-Nicolson scheme $\psi(dt)=\psi(0)+\frac{dt}{2i\hbar}H_{TB}\left(\psi(0)+\psi(dt)\right).$ The subsequent time steps are calculated with the
explicit Askar-Cakmack scheme $\psi(t+dt)=\psi(t-dt)+\frac{2dt}{i\hbar}H_{TB}\psi(t)$.
In presence of the external magnetic field the eigenstates of (3,4) need to be calculated numerically \cite{sweep}.
However, for the discussed range of the magnetic field ($B <1$ T) and the applied narrow flip area ($\lambda=4$ nm) no significant difference between the numerical eigenstates and formula of Eq. (6) used for the initial condition were found
in the wave packet evolution.

\section{Results}
\subsection{Wave packet motion}
We first consider a square flake $(x,y)\in[-300$ nm$,300$ nm$]\times [-300$ nm$,300$ nm$]$ 
with a zigzag edge at $y=\pm 300$ nm and and armchair edge at $x=\pm 300$ nm.
The packet is set in the $K'$ valley  ($\eta=-1$)
to make it move upwards (to increasing $y$ values -- see Fig. 1(b)).
Figure 2(a) shows the cross section of the packet along the $x=0$ line.
The packet indeed moves up in a stable form until it reaches the zigzag
edge of the flake.  

The same result -- as long as the packets does not reach the edge -- is obtained for the solution of the time dynamics with the continuum Hamiltonian [by Eq.(2)]. 
The absence of Zitterbewegung \cite{ci2,citer,frolo,naturd} for the wave packet that follows the Dirac equation calls for a comment. 
The velocity operator is obtained as ${\hat{v}_y}=\frac{1}{\hbar}\frac{\partial{H_\eta}}{\partial k_y}=- \eta \tau_y v_f $ \cite{citer}. For the Dirac equation this operator does not commute with the Hamiltonian  $[H_\eta,v_y]\neq 0$ which is usually invoked in the interpretation of the trembling motion \cite{citer}. However,
 the wave function localized at the
zero-line that follows the specific form used in Section II.A $\Psi=(\Psi_A, i\Psi_A)^T$ happens to be an eigenfunction of the $\hat{v}_y$ operator, with the eigenvalue $v_y=-\eta v_f$. By, the Ehrenfest theorem for we have $\frac{d}{dt}\langle \Psi |{\hat{v}_y}|\Psi\rangle=\frac{1}{i\hbar}\langle \Psi| {\hat{v}_y}H-H {\hat{v}_y}| \Psi\rangle=
\frac{1}{i\hbar}\left(\langle H \hat {v_y}\Psi|\Psi\rangle-\langle \Psi|H \hat{v_y}\Psi\rangle\right)=\frac{1}{i\hbar}\left(-\eta v_f \langle H\Psi|\Psi\rangle+\eta v_f \langle \Psi|H \Psi\rangle\right)=0$.

\subsection{Backscattering by the flake edge}
The zigzag edge of the crystal supports edge localized states. The incident packet couples to these states and in Fig. 2(a) we find formation
of a high peak of the absolute value of the wave function (see the red region at the right edge of the plot for $t\simeq$ 0.4 ps). The packet is backscattered and moves to the left with a shape restored to its original form, only in the opposite direction.

The average position $\langle y\rangle$, $\langle x\rangle$, and the size of the packet along the junction $\Delta y\equiv \sqrt{\langle (y-\langle y\rangle)^2 \rangle}$
is plotted in Fig. 3(a). The system is the same as in Fig. 2(a) only the packet
is started at $y_0=100$ nm.
In Fig. \ref{wyna}(b) we additionally plot
the square of the absolute value of the Fourier transform of the wave packet calculated
for $K$ and $K'$ valleys,
\begin{equation}
\psi({\bf k},t)=\frac{1}{2\pi} \int \psi(x,y,t) \exp(-i{\bf k}\cdot {\bf r}) {\bf dr},
\end{equation}
for ${\bf k}={\bf K}_{-1}$ (i.e. the $K'$ valley) and ${\bf k}={\bf K}_1$ ($K$ valley).
We find that as the packet reaches the edge of the flake it is scattered to the other valley
that makes it travel in the opposite direction. %The rapid oscillation of the absolute value
%of the wave function that can be spotted in Fig. 2(a) for $y>150$ nm and $t\simeq 0.4$ ps  results from the superposition of the states for both the valleys.
In Fig. 3(a) we can also see that the average $\langle x \rangle$
oscillates with a small magnitude which results from the valley flips 
at the backscattering by the edge which 
changes the sign of the exponent with $t_{SO}$ in the wave function ($\sigma_z$ is set to 1).

We find that the restoration of the soliton upon scattering is observed for both the armchair 
and the zigzag termination of the flake. 
For backscattering by the armchair edge  we flipped the potential by $-\pi/2$ angle. The packet which is still started in the $K'$ valley moves toward larger $x$ values
until it reaches the edge [Fig. 2(b)]. Here no peak of the wave function at the edge
is observed since the armchair termination does not support the edge localized states \cite{Wakabayashi2010}. 
%At the backscattering the rapid oscillation of the absolute value of the wave function due to the superposition of the plane waves with wave vectors of opposite valleys is 
%more pronounced than for the zigzag termination [cf. Fig.2(a) and (b)].

\begin{figure}
\includegraphics[width=.5\columnwidth]{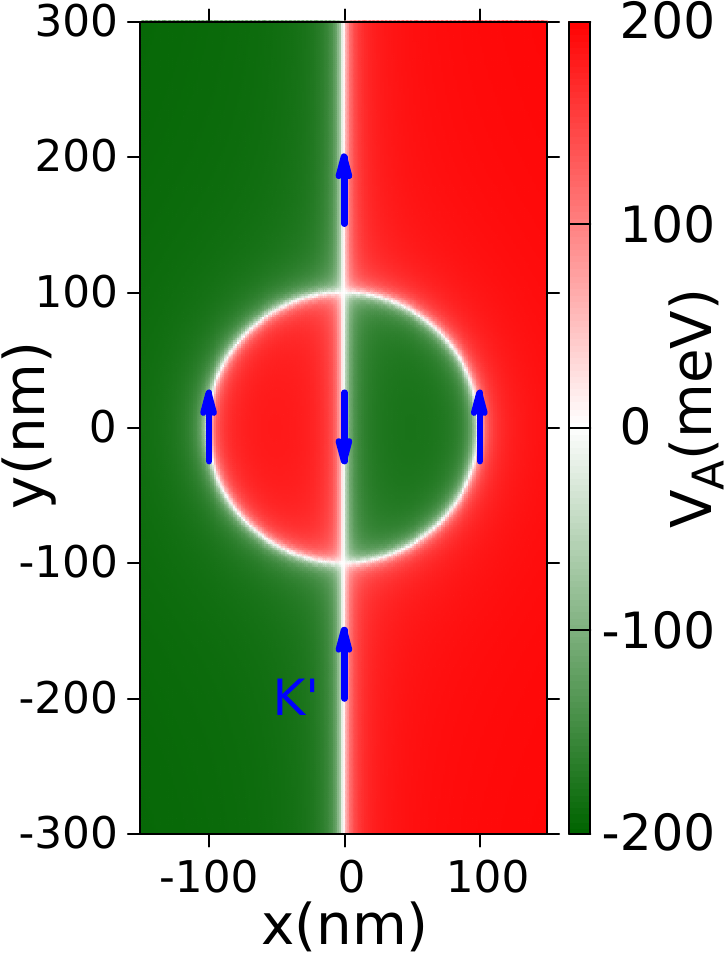}
\caption{The potential at the A sublattice for a quantum ring of radius 100 nm.
The potential at the B sublattice is assumed opposite. The arrows indicate the 
orientation of the $K'$ currents. The currents in the $K$ valley flow in the opposite direction.} \label{ring}
\end{figure}

\begin{figure*}
\begin{tabular}{lllllll}
\includegraphics[width=0.24\columnwidth]{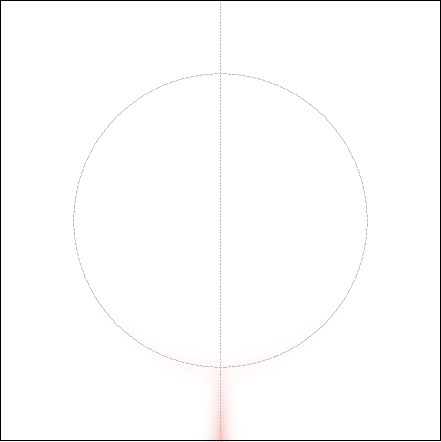}\put(-58,51){0.084ps} \put(-75,5){(a)} \put(-85,30){\rotatebox{90}{$B=0$}}&
\includegraphics[width=0.24\columnwidth]{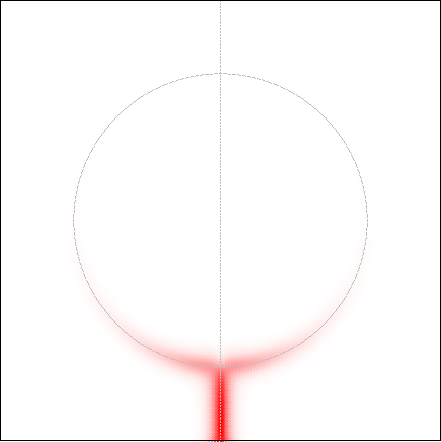}\put(-58,51){0.204ps}&
\includegraphics[width=0.24\columnwidth]{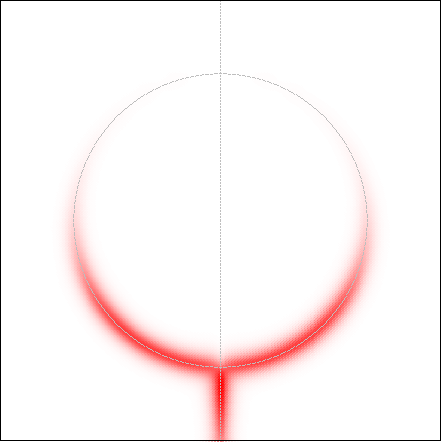}\put(-58,51){0.326}&
\includegraphics[width=0.24\columnwidth]{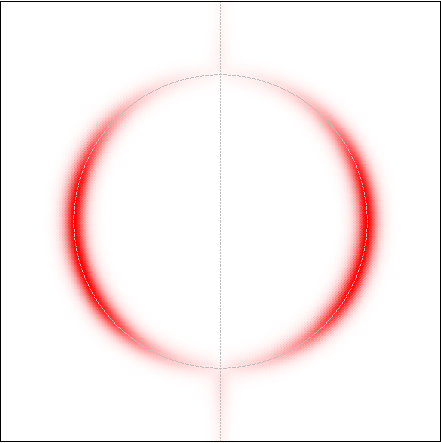}\put(-58,51){0.447}&
\includegraphics[width=0.24\columnwidth]{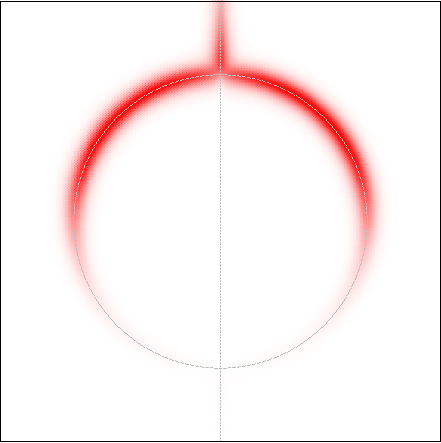}\put(-58,51){0.568}&
\includegraphics[width=0.24\columnwidth]{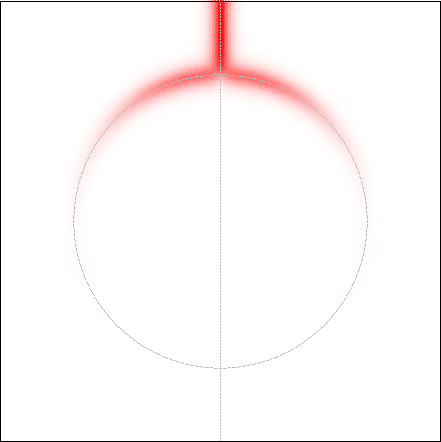}\put(-58,51){0.689}&
\includegraphics[width=0.24\columnwidth]{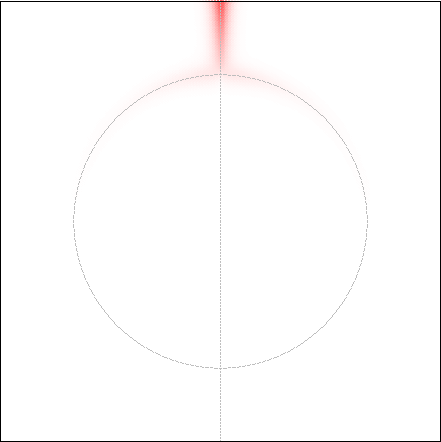}\put(-58,51){0.81} \\
\includegraphics[width=0.24\columnwidth, trim= 00 0 00 0,clip]{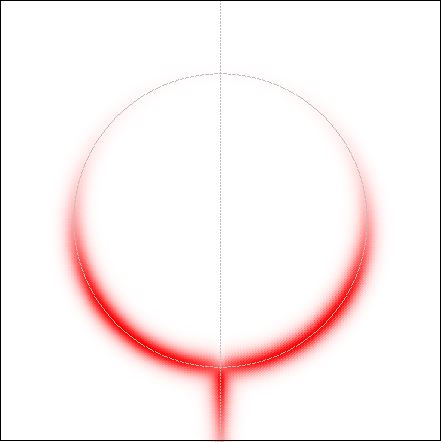}\put(-58,51){0.36ps}
\put(-75,5){(b)} \put(-85,10){\rotatebox{90}{$B=0.066$T}}&
\includegraphics[width=0.24\columnwidth, trim= 00 0 00 0,clip]{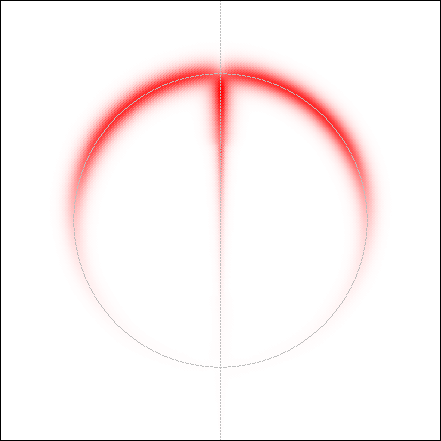}\put(-58,51){0.6ps}&
\includegraphics[width=0.24\columnwidth, trim= 00 0 00 0,clip]{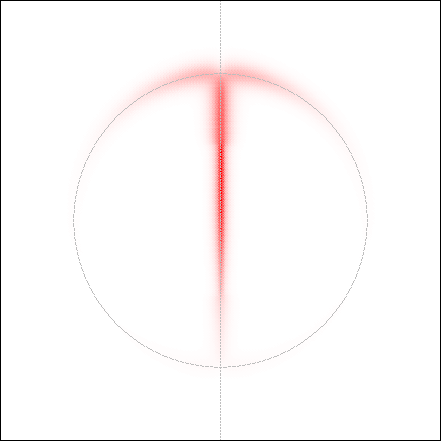}\put(-58,51){0.72}&
\includegraphics[width=0.24\columnwidth, trim= 00 0 00 0,clip]{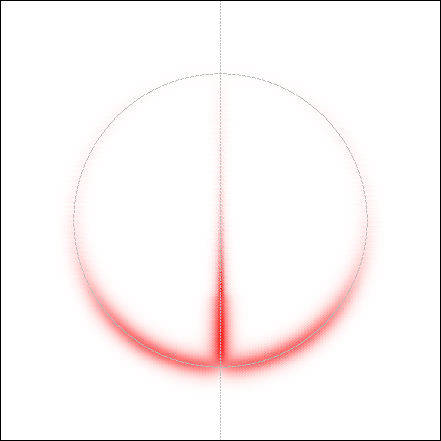}\put(-58,51){0.924}&
\includegraphics[width=0.24\columnwidth, trim= 00 0 00 0,clip]{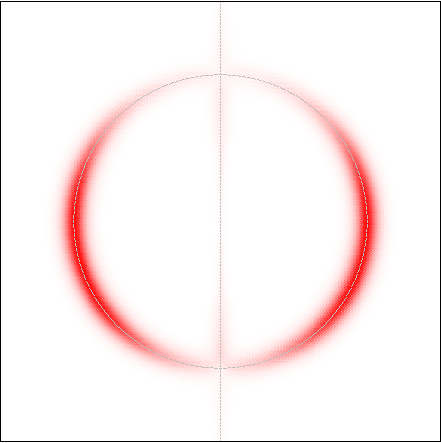}\put(-58,51){1.056}&
\includegraphics[width=0.24\columnwidth, trim= 00 0 00 0,clip]{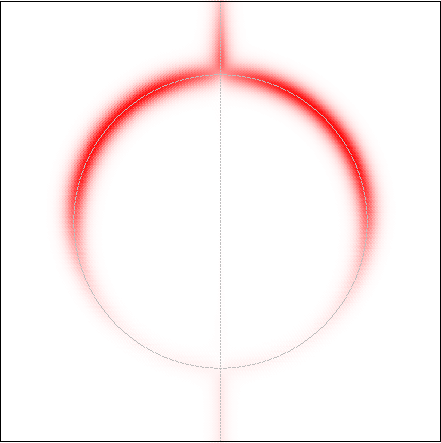}\put(-58,51){1.176}&
\includegraphics[width=0.24\columnwidth, trim= 00 0 00 0,clip]{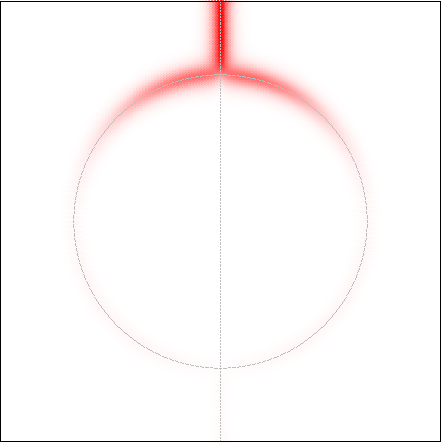}\put(-58,51){1.296} \\

\end{tabular}
\caption{The snapshots of the time evolution of the wave packet for $B=0$ (a) and $B=0.066$ T (b).
The magnetic field used in (b) corresponds to half the flux quantum threading the ring of radius 100 nm.
The time from the start of the simulation is given in picoseconds in the frames.
} \label{snap}
\end{figure*}

\begin{figure}
\begin{tabular}{l}
\includegraphics[width=0.7\columnwidth]{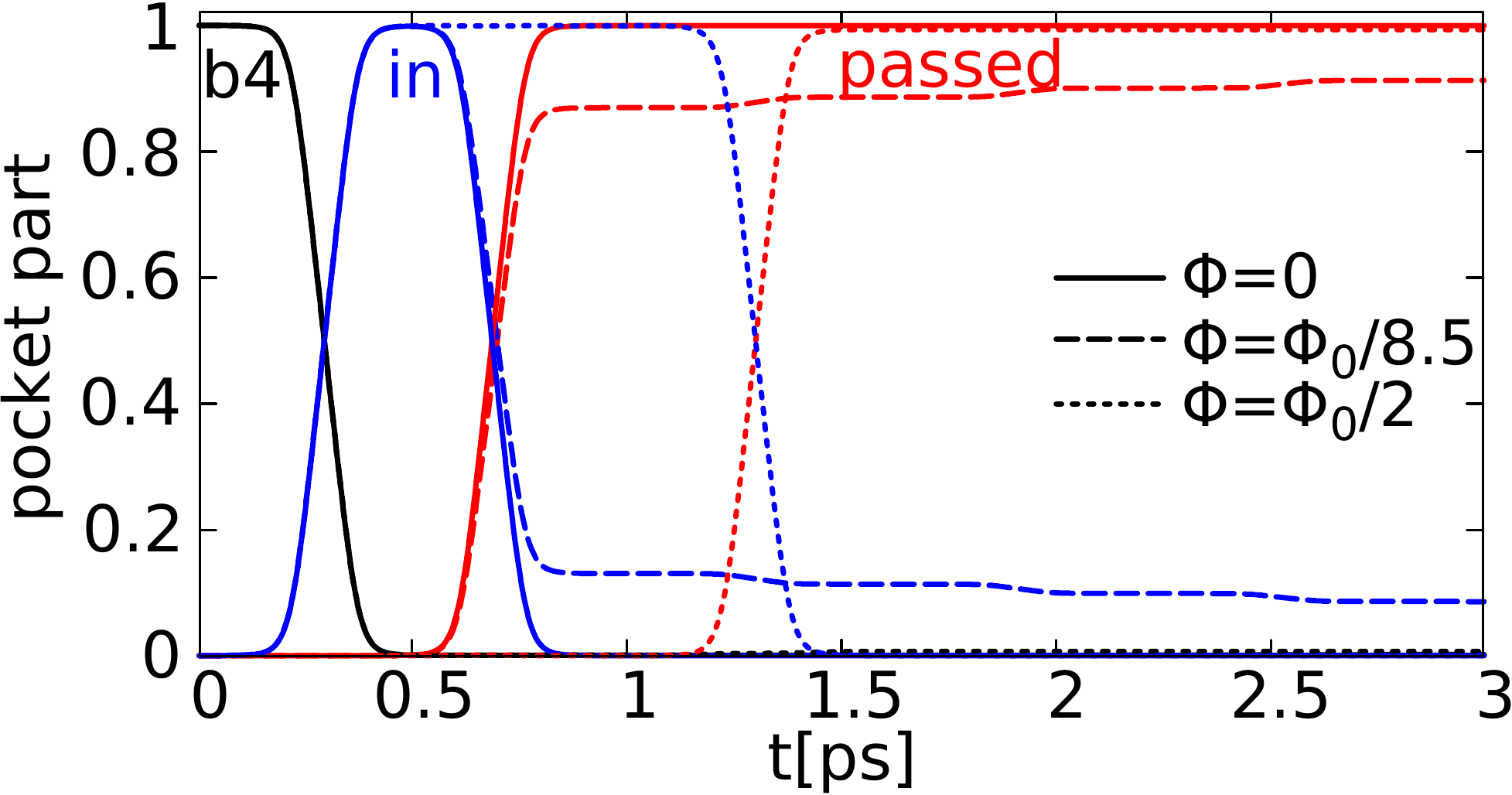}\put(1,100){(a)} \\
\includegraphics[width=0.7\columnwidth]{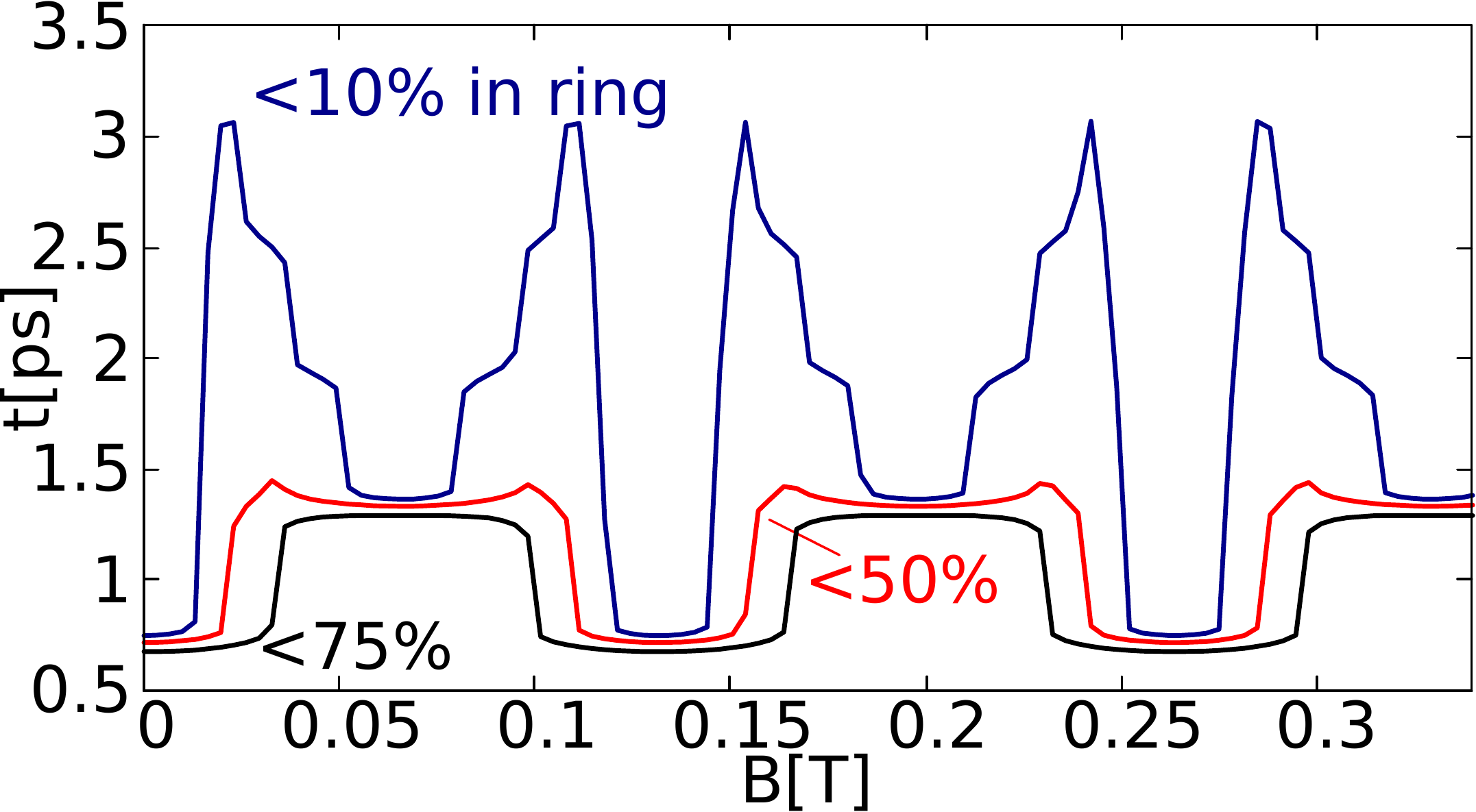}\put(1,100){(b)}\\
\includegraphics[width=0.7\columnwidth]{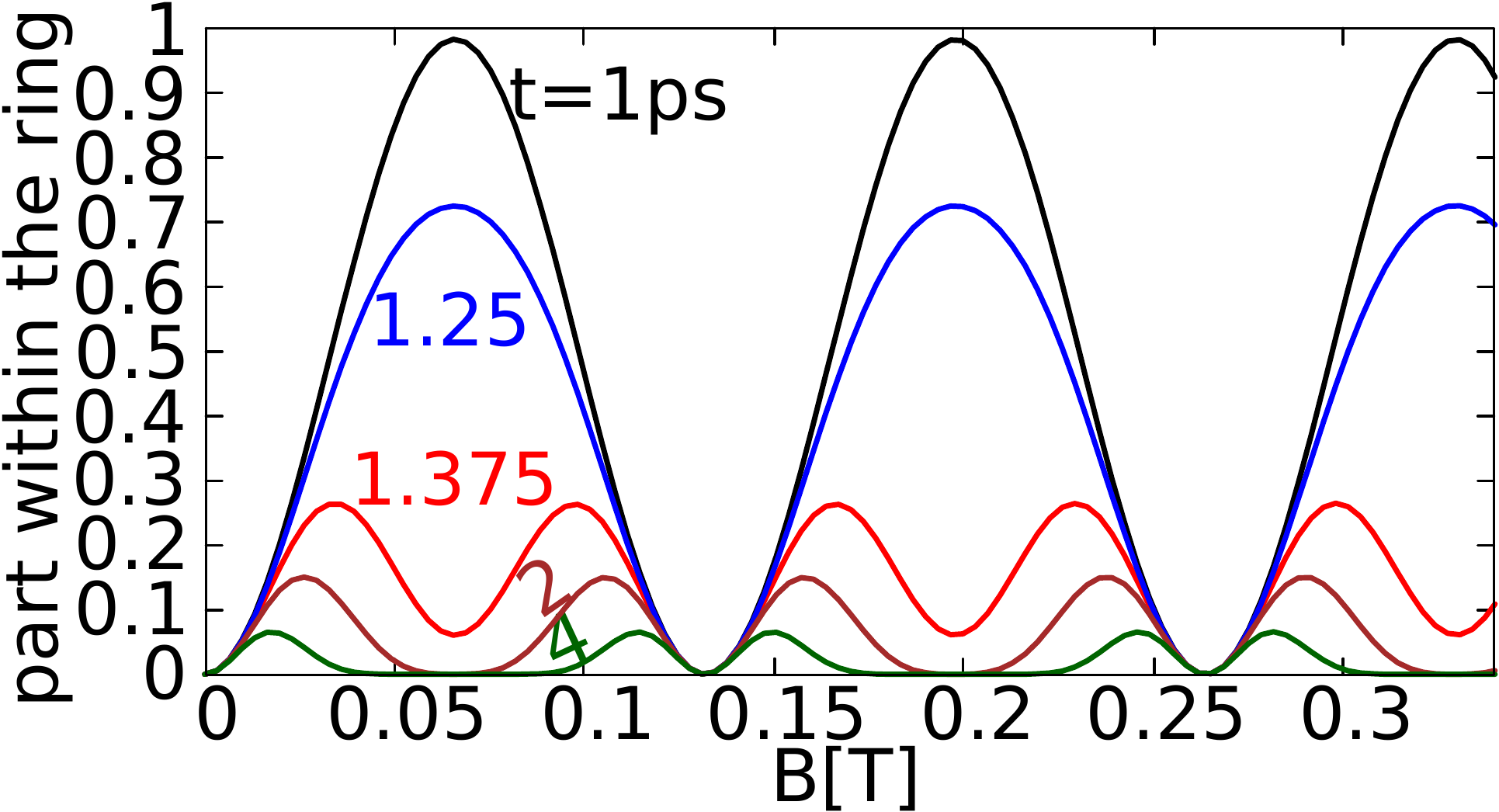}\put(1,100){(c)}
\end{tabular}
\caption{(a) The part of the wave packet before the ring (black lines),
inside the ring (blue lines) and above the ring (red lines) 
for zero magnetic field (solid lines), half of the flux quantum ($B=0.066$T) 
and the magnetic field corresponding to the flux equal to $\Phi_0/8.5$ ($B=0.0155$ T).
(b) The time at which less than 10\% (blue line), 50\% (red line), and less than 75\% of the electron density stays within the ring as a function of the external magnetic field. Local extrema	 of all the three
times are found for integer and half the flux quanta which correspond to even and odd multiples
of $0.066$ T, respectively.
(c) The part of the wave packet inside the ring for the selected moments in time from the start of the simulation.
} \label{inzain}
\end{figure}

%For inverted potential $V_g\rightarrow -V_g$ the acceptable (normalizable) localized solution is %obtained
%for $\Psi_A=i\Psi_B$, for which the chirality of the linear band is inversed $E=\eta\hbar k_y %v_F$ -- with the $K'$ ($K$) valley current flowing down (up) the band flip line. For a fixed gate structure one can redirect the valley flow by changing the sign of the potential flip. 

\begin{figure}
\includegraphics[width=0.9\columnwidth]{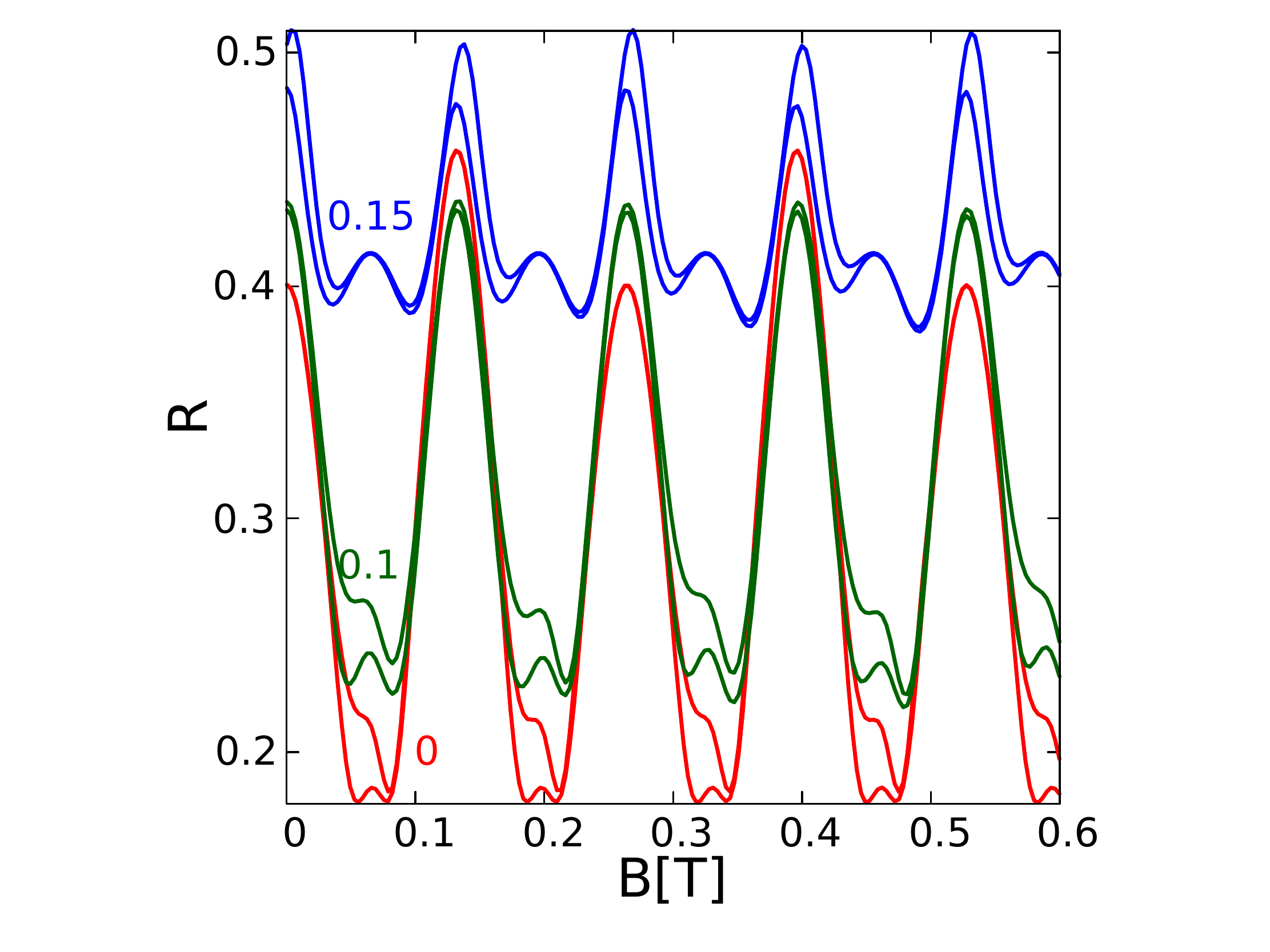}
\caption{The backscattering probability for a sharp perturbation to the potential
given by Eq. (13). The red, green  and blue lines correspond to $k_y=0$, $0.1$/nm and $0.15$/nm, respectively. For each value of $k_y$ two lines are plotted: the lower and upper bound to the
backscattering probability. The spacing between these two lines is given by the probability density
localized within the ring at the end of the simulation (6.1 ps from the start).  } \label{R}
\end{figure}

\begin{figure*}
\begin{tabular}{llllll}
\includegraphics[width=0.24\columnwidth]{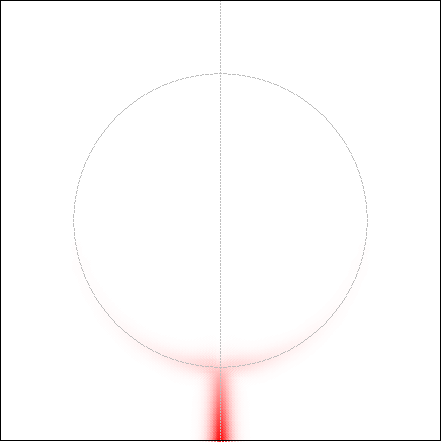}\put(-58,51){0.145ps} &
\includegraphics[width=0.24\columnwidth]{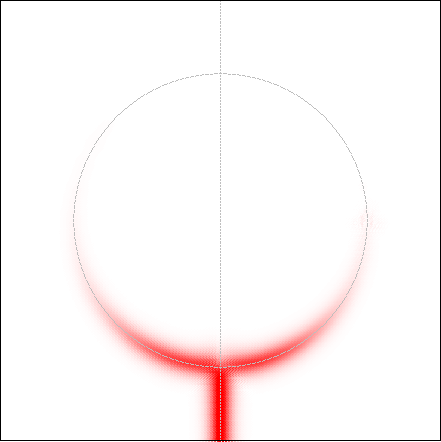}\put(-58,51){0.266ps}\put(-42,7){$\leftarrow$}\put(-28,7){$\rightarrow$}\put(-42,.5) {\small K'}&
\includegraphics[width=0.24\columnwidth]{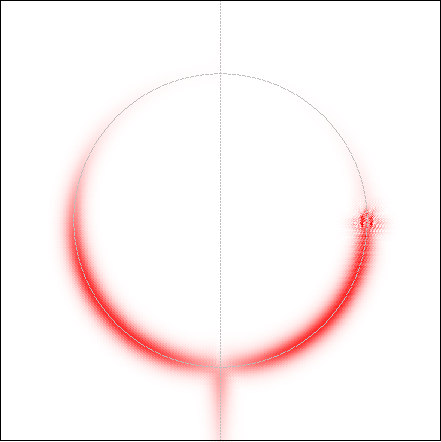}\put(-58,51){0.386}
\put(-52,12){$\nwarrow$}\put(-52,1) {\small K'}\put(-24,17){$\nearrow$}
&
\includegraphics[width=0.24\columnwidth]{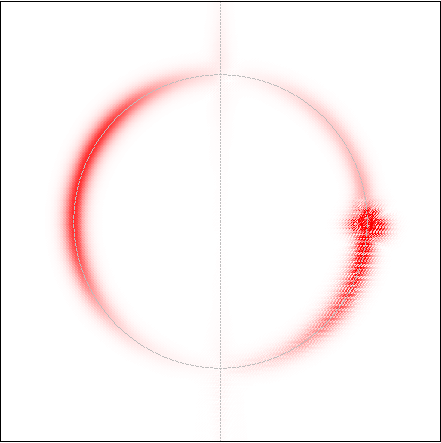}\put(-58,51){0.508}
\put(-56,26){$\uparrow$}\put(-58,16) {\small K'}\put(-21,45){$\nwarrow$}
\put(-24,17){$\nearrow$} \color{teal} \put(-14,13){$\swarrow$} \put(-10,4) {K} &
\includegraphics[width=0.24\columnwidth]{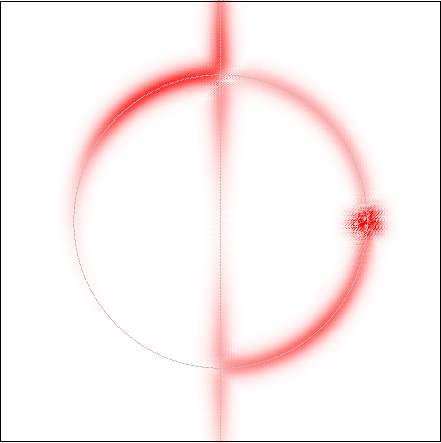}\put(-58,51){0.628}
\put(-44,39){$\nearrow$}\put(-21,45){$\nwarrow$}
\put(-29,52){$\uparrow$}
\put(-29,42){$\downarrow$}
%\put(-24,17){$\nearrow$}
 \color{teal} \put(-14,13){$\swarrow$} \put(-10,4) {K}
\put(-29,3){$\downarrow$}
\put(-29,15){$\uparrow$}
&
\includegraphics[width=0.24\columnwidth]{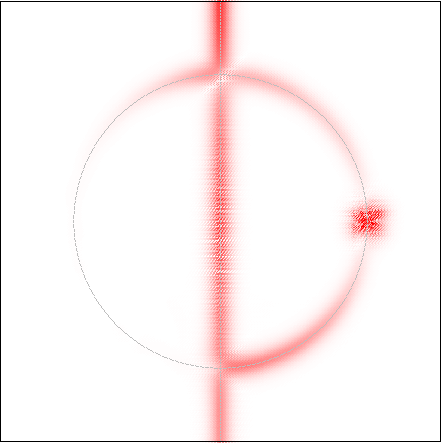}\put(-58,51){0.749}

\put(-29,52){$\uparrow$}
\put(-29,35){$\downarrow$}
%\put(-24,17){$\nearrow$}
 \color{teal} \put(-24,5){$\leftarrow$} 
\put(-29,3){$\downarrow$}
\put(-29,22){$\uparrow$}

\\
\includegraphics[width=0.24\columnwidth]{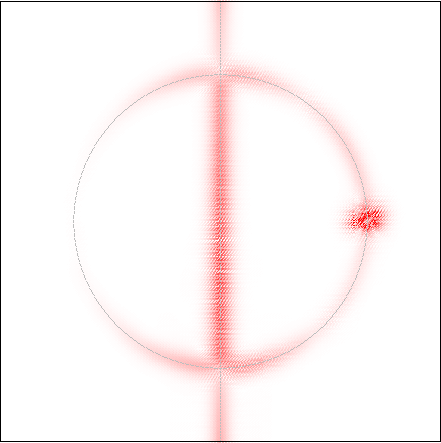}\put(-58,51){0.870} 

%\put(-42,7){$\leftarrow$}\put(-27,7){$\rightarrow$}
 \put(-29,35){$\downarrow$}\put(-29,52){$\uparrow$}\color{teal} \put(-30,3){$\downarrow$}\put(-29,22){$\uparrow$}
&
\includegraphics[width=0.24\columnwidth, trim= 00 0 00 0,clip]{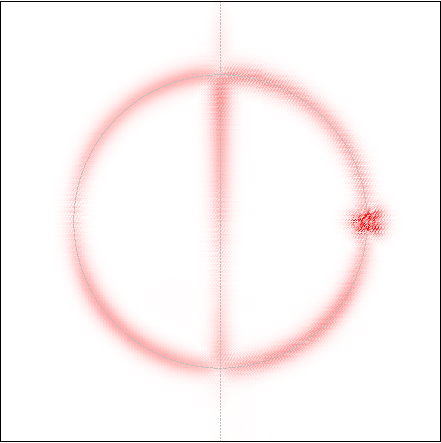}\put(-58,51){0.991ps}

\put(-42,7){$\leftarrow$}\put(-27,7){$\rightarrow$} 
\color{teal} %\put(-30,3){$\downarrow$}
\put(-29,37){$\uparrow$}
\put(-20,45){$\searrow$}
\put(-45,38){$\swarrow$}
&
\includegraphics[width=0.24\columnwidth, trim= 00 0 00 0,clip]{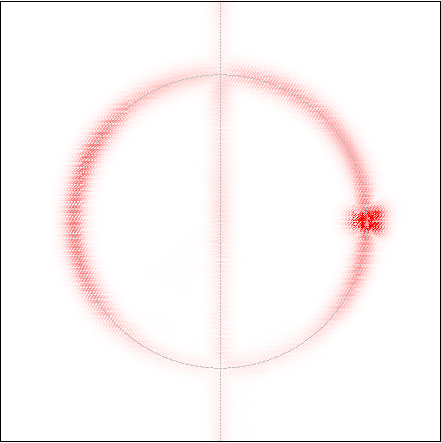}\put(-58,51){1.11ps}
\put(-56,26){$\uparrow$}\put(-6,26){$\uparrow$}
\color{teal}\put(-47,26){$\downarrow$}\put(-19,26){$\downarrow$}
&
\includegraphics[width=0.24\columnwidth, trim= 00 0 00 0,clip]{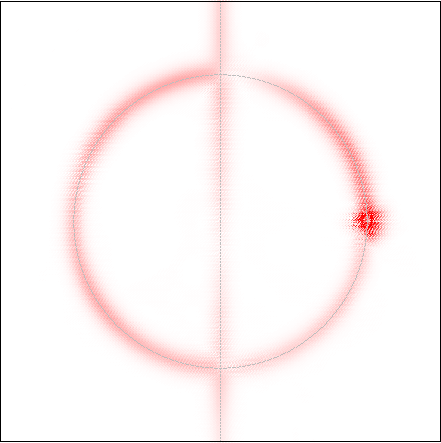}\put(-58,51){1.23}
\put(-44,39){$\nearrow$}\put(-18,42){$\nwarrow$}\put(-29,52){$\uparrow$} \put(-29,42){$\downarrow$}\
\color{teal}\put(-52,12){$\searrow$} \put(-30,3){$\downarrow$}
&
\includegraphics[width=0.24\columnwidth, trim= 00 0 00 0,clip]{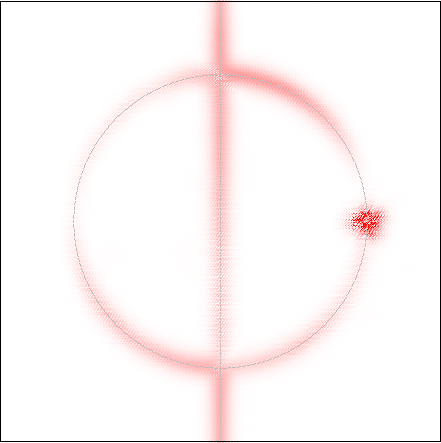}\put(-58,51){1.35}
\put(-29,52){$\uparrow$} \put(-29,42){$\downarrow$}\put(-18,42){$\nwarrow$}
\color{teal}\put(-52,12){$\searrow$}\put(-30,3){$\downarrow$}
&
\includegraphics[width=0.24\columnwidth, trim= 00 0 00 0,clip]{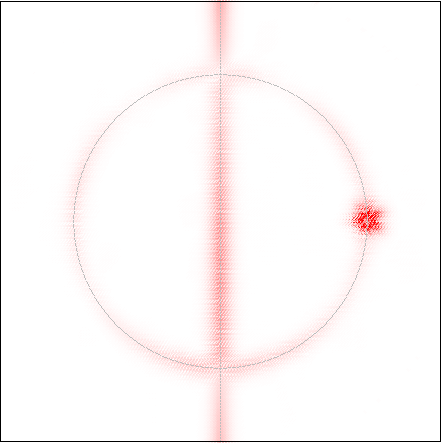}\put(-58,51){1.47}

\put(-29,35){$\downarrow$}\put(-29,52){$\uparrow$}
%\put(-24,17){$\nearrow$}
 \color{teal}
\put(-29,3){$\downarrow$}
\put(-29,22){$\uparrow$}
\end{tabular}
\caption{Snapshots of the electron density for $B=0$, $k_y=0.1$/nm (the green lines in Fig. 7(a)).
The blue (green) arrows indicate the motion of the parts of the packet moving 
in the $K (K')$ valley. The time from the start of the simulation is given in picoseconds in the frames. } \label{wiring}
\end{figure*}
\subsection{Quantum ring}
A form of a quantum ring with the chiral waveguides for the electron flow can be defined with an  engineering of the electric field. For that purpose
one needs a local inversion of the electric field  which requires additional top and bottom gates in the system. Here we consider a circular area of radius $R$,
and set the model potential at the $A$ sublattice to
\begin{align}
V_A=  \frac{4V_g}{\pi^2}\arctan \left(\frac{r-R}{\lambda}\right)\arctan \left(\frac{x}{\lambda}\right),
\end{align}
where $R=100$ nm is the circle radius with the center at the origin and $r=\sqrt{x^2+y^2}$ is the distance from the origin. The potential at the $A$ sublattice is plotted in Fig. \ref{ring}. As above we take $V_B({\bf r})=-V_A({\bf r})$. 
Fig. \ref{ring} shows also the direction of the flow for currents in the $K'$ valley -- with 
the negative (positive) potential at left (right) side of the current flow. 
When the line of $V_A=0$ meets the ring at $y=-100$ nm the $K'$ current can go to either
the left or the right arm of the ring. This potential profile introduces  a beam splitter
for the electron wave packet in this way.  The central bar of the ring is inaccessible for the $K'$ current going up.

In quantum rings \cite{qr} Aharonov-Bohm \cite{ab} oscillations of coherent conductance
are observed that in the Landauer-B\"uttiker \cite{labu} approach are explained as due to variation of the electron transfer probability 
across the ring with phase shifts accumulated from the vector potential of the magnetic field.
In the present system the electron backscattering from the ring is prohibited by the topological protection of the  valley current, so the transfer probability of the soliton packet is 1 for any magnetic field. 
However, we find that the time that the electron spends within the ring
changes due to the phase shifts introduced by the vector potential.

In this section and in the rest of the paper we neglect the intrinsic spin-orbit coupling ($t_{SO}=0$) that introduces a weak asymmetry in the electron injection to the ring due to the spin-valley dependent shift off the zero line [cf. Fig. 3(a) and Eq. (6)]. { The calculations for the quantum ring
require long leads to prevent return of the packet to the scattering area upon reflection
from the edge of the crystal. For long wave packets the entire computational box is taken long up to 6 $\mu$m. Systems these long are treated with the scaling method of Ref. \cite{scaling} 
for which the crystal constant is scaled as $a_s=a s_f$ with the hopping parameter $t_s=t/s_f$.
In Hamiltonian (8) $t_s$ replaces $t$, and $a_s$ replaces $a$ while the Si ions are generated in the computational box. We use the scaling factor $s_f=3$ or $4$ in the calculations for silicene below. }

Figure 5(a) shows the snapshots of the simulation of the packet transfer across the ring
for $B=0$. For $t=0$ the packet is started 350 nm below the center of the ring of radius 100 nm.
The snapshots taken at 0.204 ps and 0.325 ps show that the packet is split into two parts
at the entrance to the ring.
In both the left and right arms of the ring the $K'$ current moves up leaving the 
negative potential at the A sublattice at the left-hand side [see Fig. 4].
The split packets meet at the exit of the ring [$t=0.568$ ps and $t=0.689$ ps]
with the same phase and the packet of its original size is restored [$t=0.810$ ps]. 

Figure 6(a) shows the parts of the packet before ("b4", black lines) the ring, 
within ("in", blue lines), as well as the transferred part ("passed", red lines) as a function
of time. The results for $B=0$ that correspond to Fig. 5(a) are plotted with the
solid lines. For $t=0.75$ ps the entire packet is transferred above the ring.

In Fig. 5(b) we plotted the snapshots of the simulation for $B=0.066$ T
which corresponds to half the flux quantum $B\pi R^2=\Phi_0/2=\frac{h}{2e}$ threading the ring with $R=100$nm. 
The parts of the packet that meet at the exit ($t=0.6$ ps) acquire a relative $\pi$ phase 
due to the  Aharonov-Bohm effect.
A node of the wave function is formed at the exit to the upper channel.
A nondestructive interference  is observed within the area  below the exit from the ring,
which directs the packet to the internal bar within the ring ($t=0.72$ps), i.e. to the only path where the $K'$ current can go for the exit to the upper channel blocked by the Aharonov-Bohm effect. The bar is transparent for the $K'$ current going down [Fig.4]. The packet is split again to the left and right arms of the ring at $0.924$ ps.   
After the second round the parts of the packet meet in phase $2\pi$ at the exit and the packet
smoothly leaves the ring [$t=1.176$ ps, $t=1.296$ ps].

In Fig. 6(a) the results for half the flux quantum are plotted with the dotted line.
The packet is transferred to the exit with a delay but completely and in a single move. 
For comparison in Fig. 6(a) the results for the magnetic field of $0.0155$ T
 which corresponds to $\Phi_0/8.5$ are plotted  with the dashed lines. Here a part of the packet passes to the 
exit as fast as for $B=0$, but due to a phase difference 
at the exit a part of the packet stays inside the ring and leaves it in portions at
subsequent rounds, which produces the steps in the dashed lines in Fig. 6(a).  

In Fig. 6(b) we plotted the moments in time for which less than 75\%, less than 50\% and
less than 10\% stays within the ring as a function of the external magnetic field.
The plot is periodic with the Aharonov-Bohm period of $0.132$ T. 
The ring is emptied the fastest for the integer and half quanta. The result of Fig. 6(a) 
for $B=0.0155$ T corresponds to a local maximum of the time for which more than 10\% of
the packet stays within the ring. 
Finally Fig. 6(c) shows the part of the packet contained within the ring for a fixed moment
in time as a function of the magnetic field. For $t=1$ps and $1.25$ps the ring-confined part is locally maximal for the magnetic field corresponding to odd multiples of half the flux quantum
$B=(2n+1)0.066$ T. For longer times these maxima are turned into minima due to 
compensation of the phase difference after the second round of electron circulation [Fig. 5(b)] within the ring.

\subsection{Intervalley scattering and conductance oscillations}
For the potential profile plotted in Fig. 4 the transfer probability can fall below 1 
only provided that a intervalley transition is present within the ring.
% otherwise no backscattering can occur.
 The intervalley scattering is introduced by potential
variation that is short on the atomistic scale. For the modeling we introduced a point potential defect
of a Lorentzian  form 
%\begin{equation} V_d({\bf r})=\frac{V_g}{1+{({\bf r}-{\bf r}_d)^2}/{(12a)^2}}, \end{equation}
{
\begin{equation} V_d({\bf r})=\frac{V_g}{1+{({\bf r}-{\bf r}_d)^2}/{l^2}}, \end{equation}
where ${\bf r}_d=(R,0,0)$, and $V_g=7.2/s_f^2$ eV and $l=12a=4.7$ nm. In the scaling approach \cite{scaling} the smooth potential that changes slowly on the atomic scale, in particular the one given by Eq. (1) and (12) stays invariant with $s_f$. For the abrupt short-range defect potential we found that scaling of $V_g$ with $s_f$ is necessary to keep the same effectiveness of the intervalley scattering.
 The defect potential $V_d$ enters with the same sign to potential on both sublattices,
as $V_A+V_d$ on  sublattice A and $V_B+V_d=-V_A+V_d$ on sublattice B. }

%For the clean ring the
%results stay very similar for any value of $k_y$ provided that it is small (i.e., $|k_y|\lesssim %0.15$/nm).
In presence of the defect, the results start to change significantly with $k_y$. 
The backscattering probability  $R$ as a function of the external field is given in Fig. 7
for $k_y=0$, $0.1$/nm and $0.15$ /nm. 
For each value of $k_y$ two lines are plotted, which are the minimal and maximal bound for the 
backscattering probability. The difference between the two is determined by the part of the electron
packet that stays within the ring at the end of the simulation (6.1 ps). 
$R$ as a function of $B$ is approximately periodic with the flux quantum.% For higher $k_y$ the oscillations
%of $R$ have lower amplitude.
 The behavior of the electron packet is displayed in Fig. 8 for $B=0$ and $k_y=0.1$/nm. The packet is incident in the $K'$ valley. For $t=0.508$ ps we spot the scattering center
near ${\bf r}_d$. A part of the packet passes across the defect moving still in $K'$ valley, and
a larger part is backscattered and move in the direction which is only allowed for the $K$ valley. 
For $t=0.628$ ps the $K$ current reaches the entrance to the ring, a part of it is backscattered
to the input channel, and another goes up along the bar. The electron packets of opposite valley
meet within the bar for $t=0.749$ ps and 0.87 ps. 
For $t=0.991$ ps we can see recycling of currents for both the valleys: the $K$ valley current cannot pass to the output channel and the $K'$ one to the input channel.
The opposite valley currents meet again, this time in the arms of the ring for $t=1.11$ ps. For $t=1.47$ ps the packet distribution
 is similar to $t=0.87$ ps, only a smaller portion of the electron packet is still
present within the ring.

\subsection{The limit of long wave packets}

The wave packet dynamics in the limit of a large size of the packet should approach the conditions of the stationary electron flow. In order to study the limit of a plane incident wave we 
consider a Gaussian envelope of the packet, i.e. in the initial condition given by Eqs. (9) and (10) we set $\chi=\exp(-(y-y_0)^2/4\sigma ^2)/(2\pi\sigma^2)^{1/4}$. The Fourier transform of the packet 
produces the probability density with the standard deviation of $\sigma_k=\frac{1}{2\sigma}$ in the wave vector space.

\begin{figure}
\begin{tabular}{l}
\includegraphics[width=0.9\columnwidth]{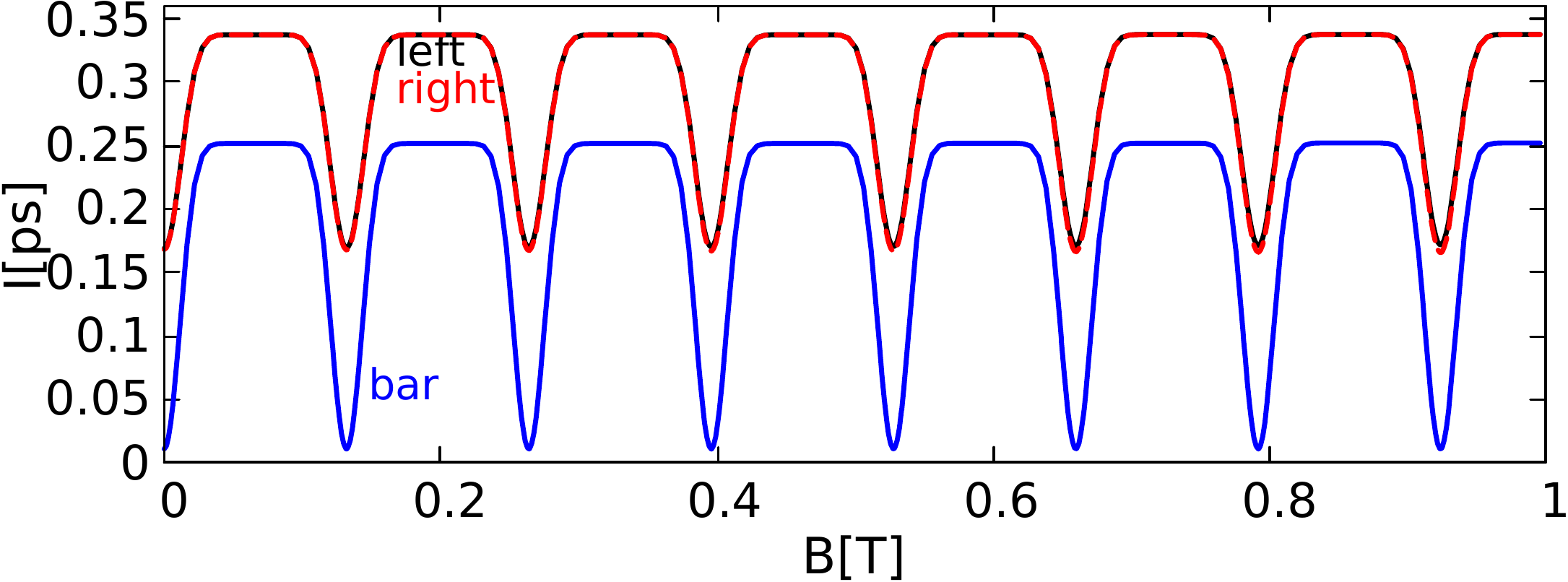}\put(1,80){(a)}\put(-200,85){$\sigma=30$ nm} \\
\includegraphics[width=0.9\columnwidth]{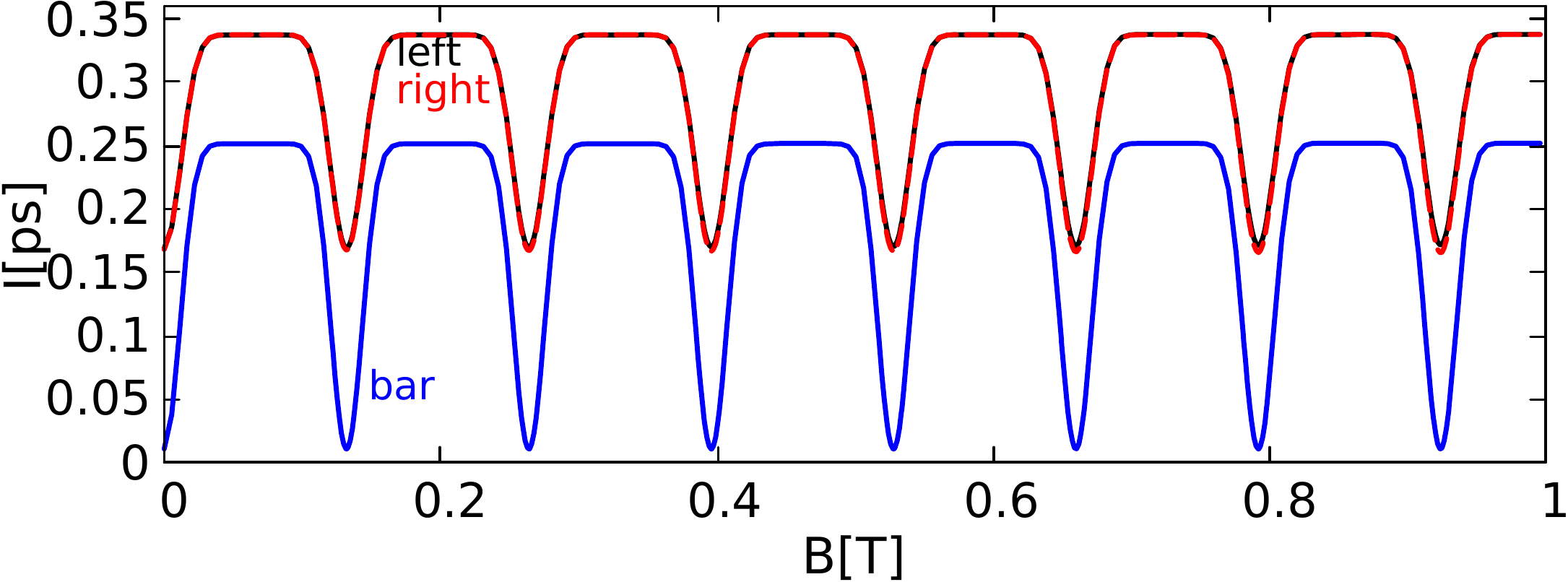}\put(1,80){(b)}\put(-200,85){$\sigma=60$ nm}\\
\includegraphics[width=0.9\columnwidth]{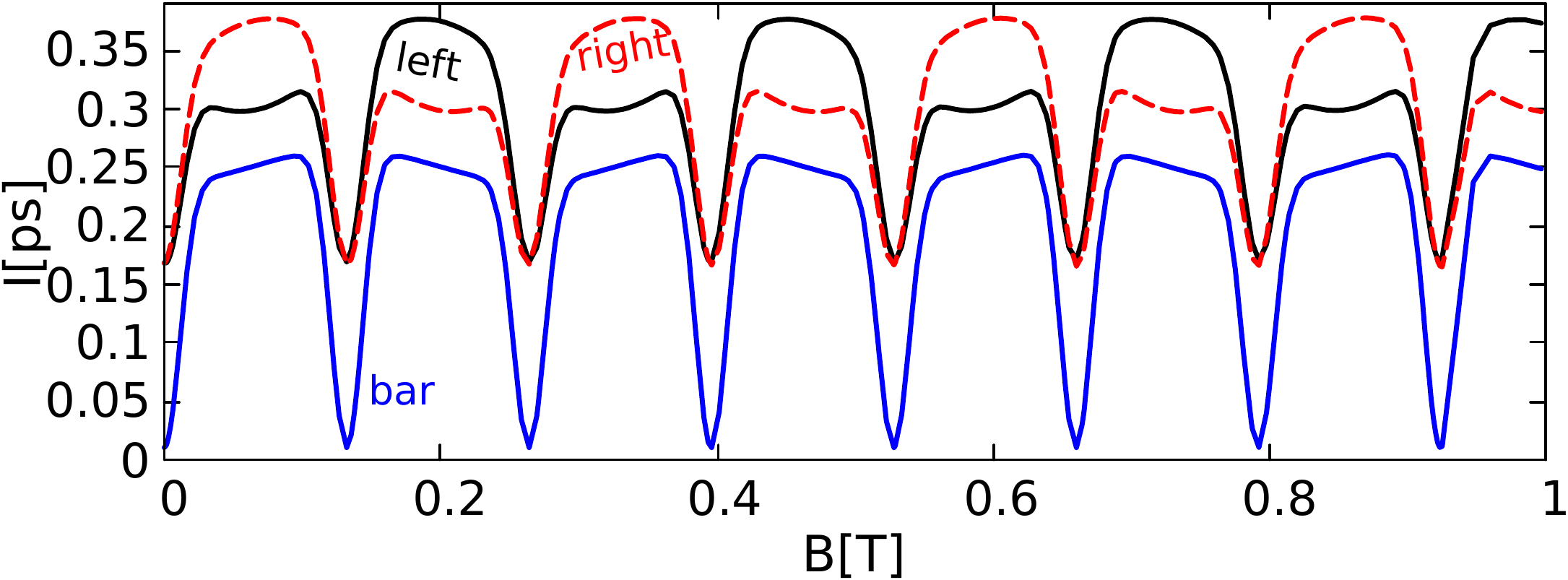}\put(1,80){(c)}\put(-200,85){$\sigma=120$ nm} \\
\includegraphics[width=0.9\columnwidth]{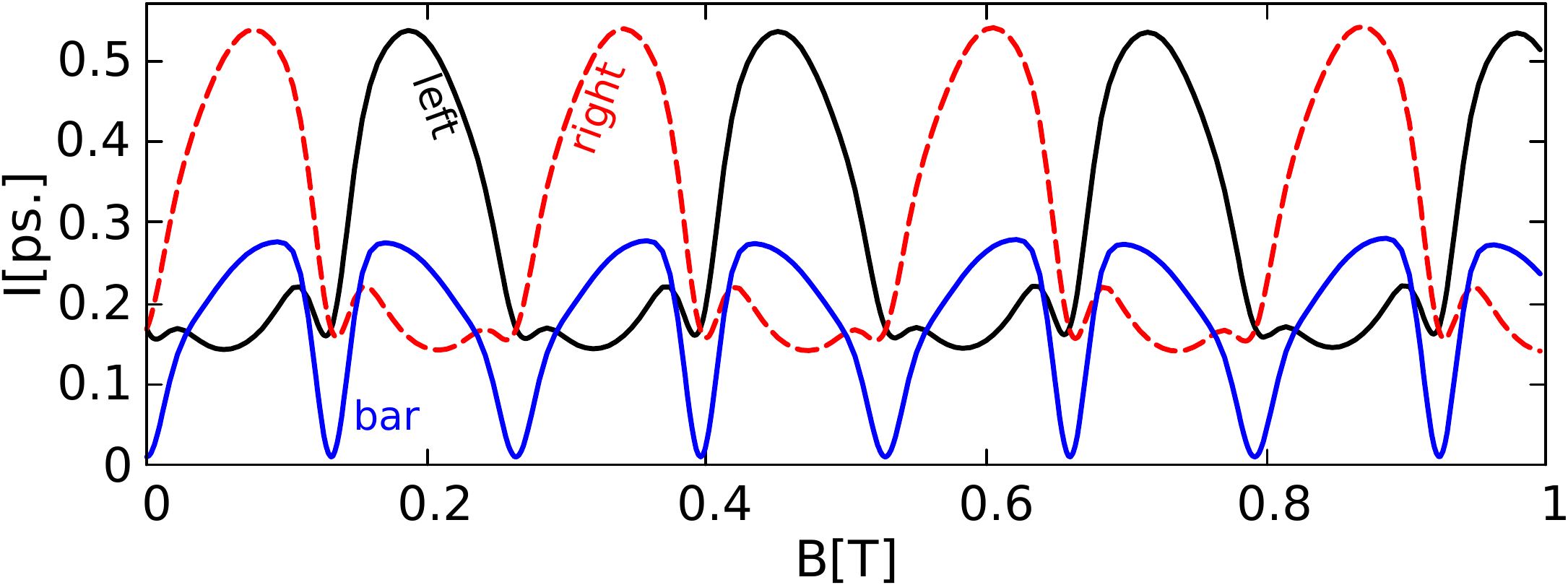}\put(1,80){(d)}\put(-200,85){$\sigma=240$ nm}
\end{tabular}
\caption {Probability density in the left (red dashed line), and right (black line) arms 
of the ring and in the central bar (blue line) integrated over time, for a Gaussian enveloppe of
the wave packet in the initial condition $\exp(-(y-y_0)^2/4\sigma ^2)$, with $\sigma$ given
in the figure. The wave vector applied in the initial condition is $k_y=0.01$/nm.
} \label{lmit} \color{black}
\end{figure}
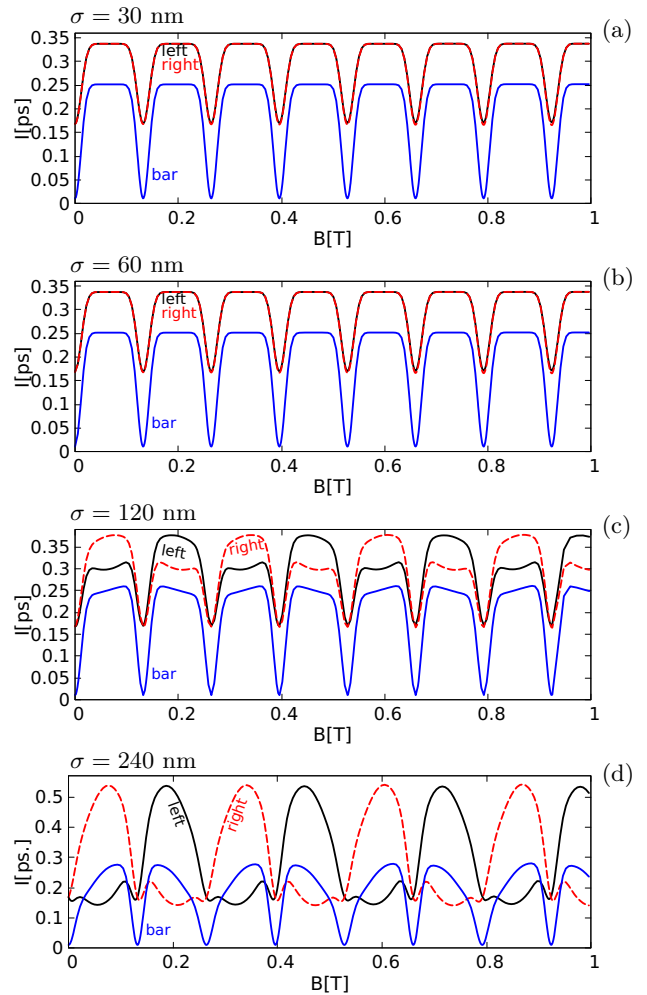

We solve  the time evolution and integrate the parts of the packet in the left and right arms
of the ring and in the center bar over time as the packet transfers the ring. 
 In Fig. \ref{lmit} we plot the results
for increasing length of the packet in the initial condition. 
The results for $\sigma=30$ nm and $\sigma=60$ nm are nearly identical. The packet passess equally
through the left and right arms of the ring. The integrals in Fig. \ref{lmit}(a) and  Fig. \ref{lmit}(b) are periodic functions of the magnetic field with the quantum of the  flux threading the ring of radius $R=100$ nm, i.e. $B=0.132$ T, as above. However, an asymmetry of the electron transfer 
across the arms appears for $\sigma=120$ nm [Fig. 9(c)] and becomes very strong for $\sigma=240$ nm [Fig. 9(d)]. Moreover, as the result of this asymmetry, the period of the integrals doubles. 
%The packets are injected to the arms of the ring symetrically for any $B$, only once
%the scattering density appears in the entire system the asymmetry is formed. 
For explanation of this effect the solution of the stationary scattering problem of the next subsection is helpful.   

\subsection{Stationary electron flow}

\begin{figure}
\begin{tabular}{l}
\includegraphics[width=0.9\columnwidth]{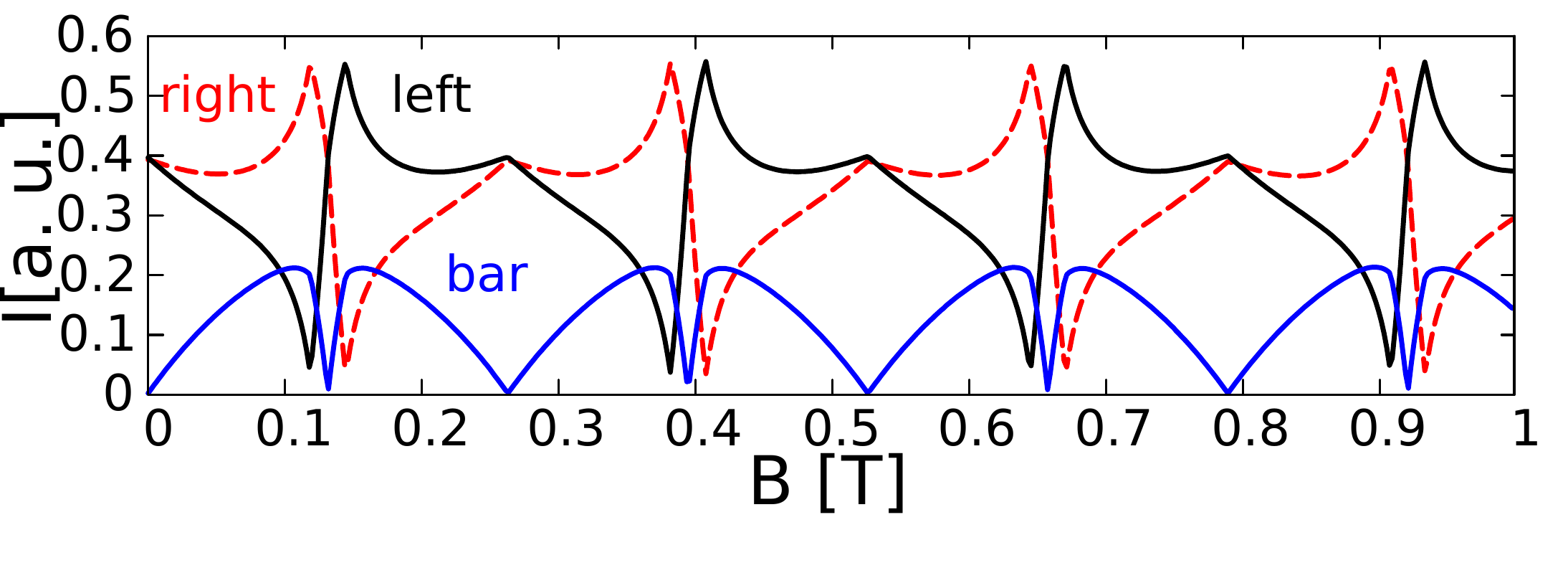}\put(1,80){(a)}\\
\includegraphics[width=0.3\columnwidth]{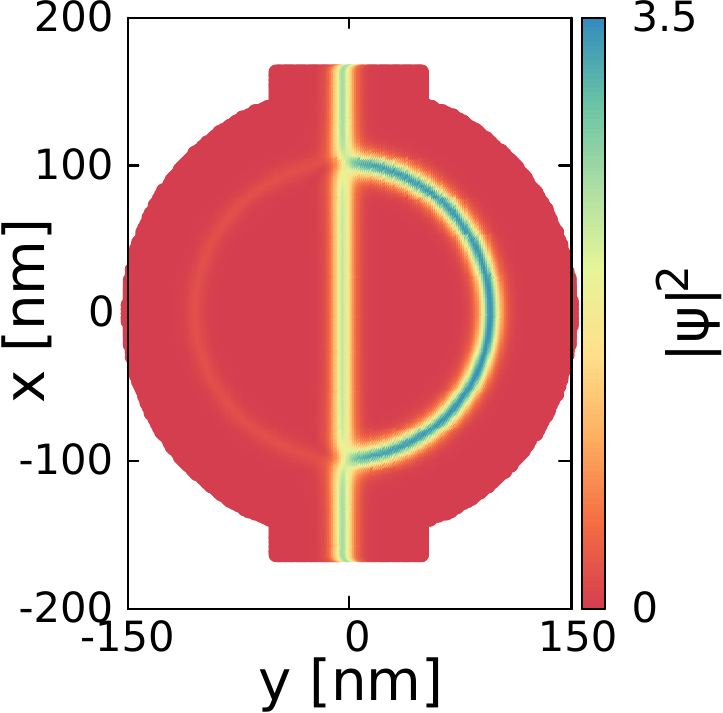} \put(-50,80){(b) 0.12 T}\;
\includegraphics[width=0.3\columnwidth]{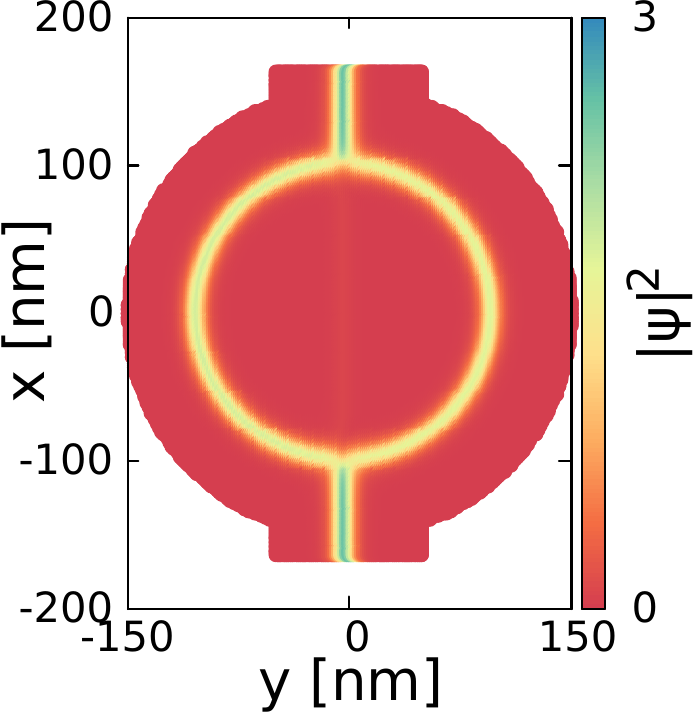} \put(-50,80){(c)  0.13 T}\;
\includegraphics[width=0.3\columnwidth]{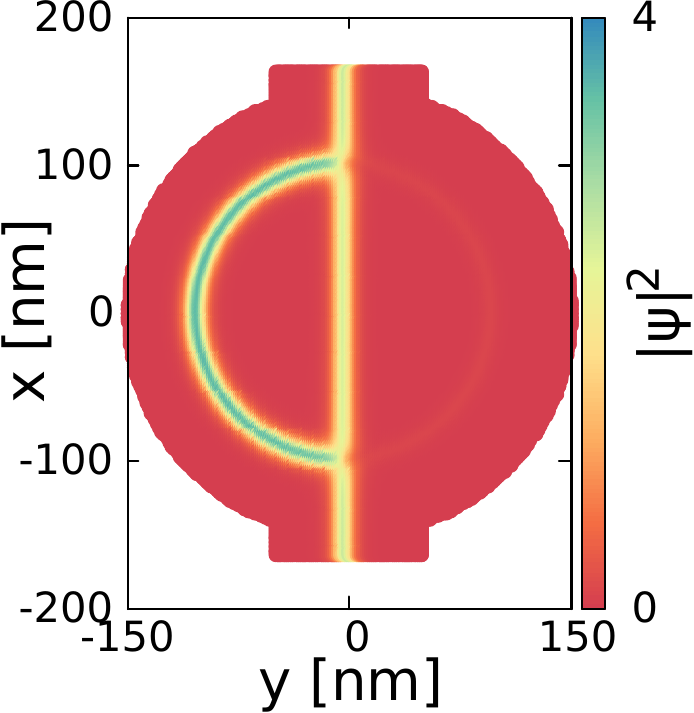} \put(-50,80){(d) 0.14 T }\\
\includegraphics[width=0.8\columnwidth]{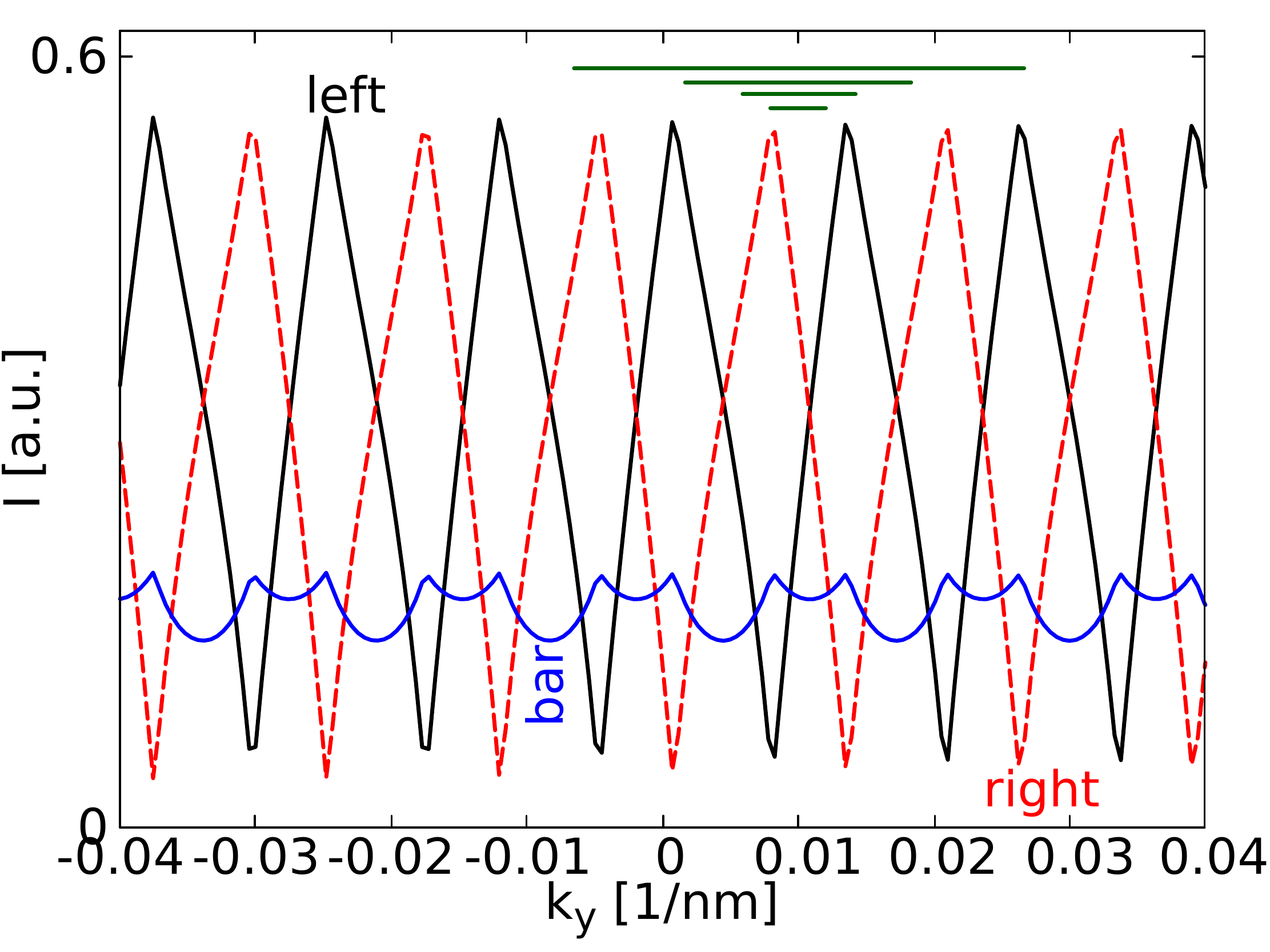}\put(1,80){(e)}\\
\includegraphics[width=0.6\columnwidth]{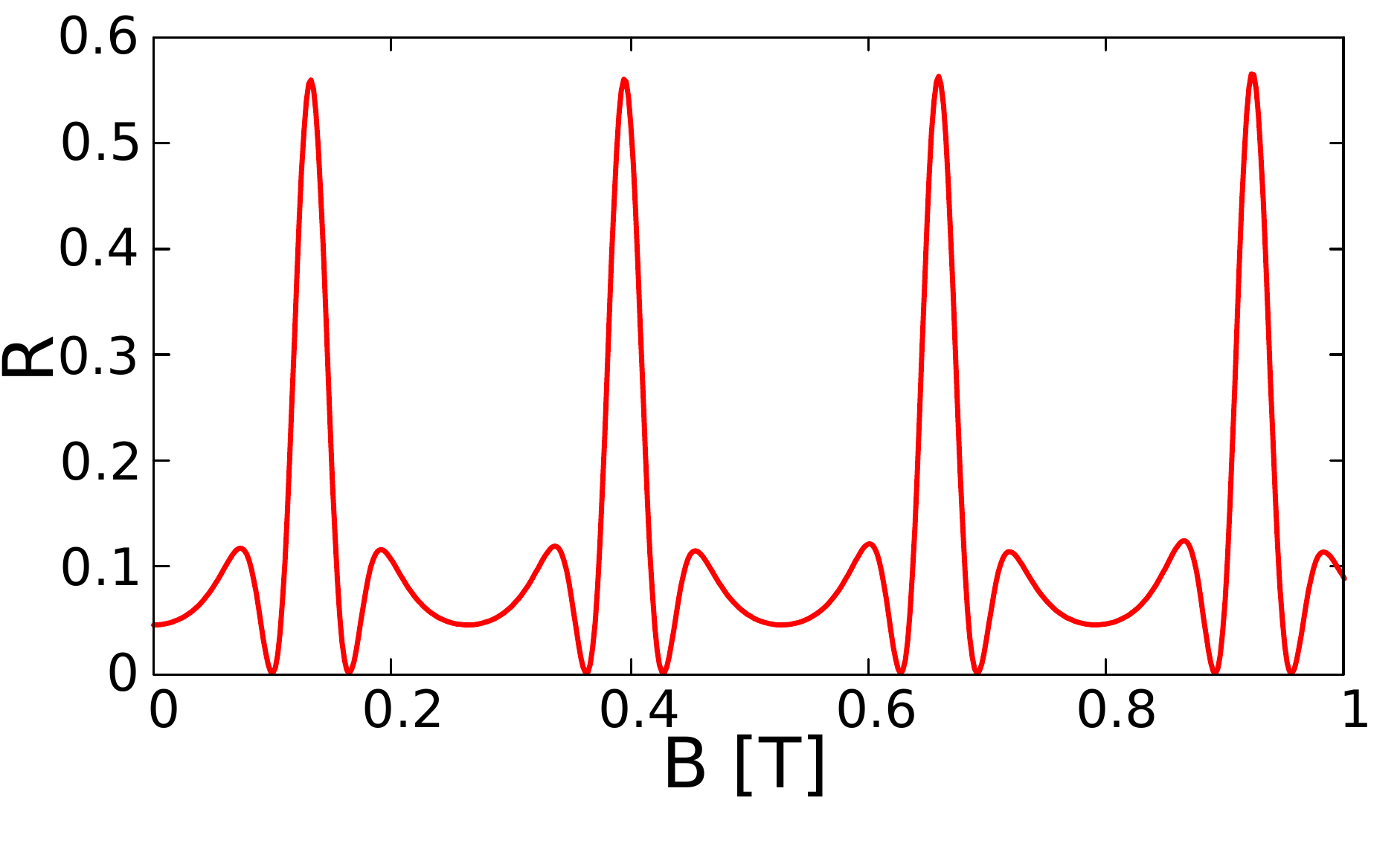}\put(1,80){(f)}\\
\end{tabular}
\caption { Solutions of the stationary scattering problem for $E_F=6.4$ meV,
which corresponds to the wave vector $k_y=0.01/$nm as calculated with respect to the $K'$ valley.
(a) Scattering  density integrated over  the left (red dashed line), the right (black line) arm
of the ring, and in the central bar (blue line).
(b-d) The scattering density for $B=0.12$, $0.13$, $0.14$ T.
(e) Same as (a) only as a function of $k_y$ for $B=0.075$T. The horizontal bars show
the segments $[k_y-\sigma_k,k_y+\sigma_k]$ with the standard deviation of the Gaussian packet
$\sigma_k=\frac{1}{2\sigma}$ for $\sigma=30$, 60, 120, and 240 nm, from top to bottom.
(f) Backscattering probability obtained for the defect potential given by Eq. (13). 
} \label{ciajm}
\end{figure}

We solve the standard stationary quantum scattering problem for the Fermi level electron
using the wave function matching technique \cite{wfm} for the atomistic tight binding Hamiltonian. We set  $E_F=6.4$ meV for which the Fermi wave vector is displaced by $k_y=0.01$/nm from the $K'$ Dirac point.% to compare the results with Fig. \ref{lmit}.
The integral over the scattering density in space is plotted in Fig. \ref{ciajm}(a).
The result corresponds very well with Fig. \ref{lmit}(d), only the features are more abrupt
in the stationary case, which is due to the presence of a finite range of $k_y$ in the wave packet dynamics. 

Figures \ref{ciajm}(b,c,d) show the scattering density for 0.12, 0.13 and 0.14 T, respectively.
For 0.13 T [Fig.\ref{ciajm}(c)] the parts of the electron density passing through both the arms of the ring meet in phase at the exit to the ring, and the electron wave function does not enter the central bar. However, for slightly different magnetic field [Fig.\ref{ciajm}(b,d)] a phase difference appears, the
part of the wave function is injected to the central bar from above, and the interference
within the ring promotes right or left arm of the ring. In the extreme conditions of Figs. \ref{ciajm}(b,d) the
 electron circulates around a {\it half } of the entire ring, which is the origin of the period doubling 
on the magnetic field scale, as the area for the magnetic field flux is halved. 

In the time-dependent dynamics for short packets the injection to both the arms of the ring is nearly ideally symmetric. Figure \ref{ciajm}(e) gives the integrals for the stationary flow
for $B=0.075$ T as a function of $k_y$.  The scattering density switches very fast from the left
to the right arm of the ring.  The horizontal bars at the top of the Figure
show the segments of $k_y$ from $0.01 \frac{1}{\mathrm{nm}} \pm \sigma_k=0.01 \frac{1}{\mathrm{nm}} \pm \frac{1}{2\sigma}$, for $\sigma=30,60,120$ and 240 nm (from top to bottom). Only for large $\sigma$ the asymmetry survives averaging over $k_y$ range contained within the packet, hence
the symmetric transfer for shorter packets. 

Figure \ref{ciajm}(f) shows the backscattering probability $R$
for the ring with the defect given by Eq. (13). The conductance $G=(1-R)\frac{2e^2}{h}$ 
has the period of $0.264$ T -- corresponding to the flux quantum 
threading half of the ring
in consistence with the period of the scattering density integrals of Fig. \ref{ciajm}(a).

\subsection{Quantum rings defined by zero lines in bilayer graphene}
Qualitatively similar results for the transfer across the quantum rings
defined by the zero lines of the symmetry breaking electric field are
found for the bilayer graphene.  

For bilayer graphene we use the atomistic tight-binding Hamiltonian spanned by $p_z$ orbitals,
\begin{equation}
   H=\sum_{\langle i,j\rangle }\left(t_{ij} c_i^\dagger c_j+h.c.\right)+\sum_i V({\bf r}_i) c_i^\dagger c_i, 
\label{eq:dh}
\end{equation}
where  $V({\bf r}_i)$ is the external potential 
at the $i$-th site at position $\mathbf{r}_i$, and in the first term we sum over the nearest neighbors. We use the tight-binding parametrization of
Bernal stacked layers \cite{Partoens}, with 
$t_{ij}=-3.12$ eV for the nearest neighbors within the same layer. 
 For the interlayer coupling, we take $t_{ij}=-0.377$ eV for the A-B dimers, $t_{ij}=-0.29$ eV for skew interlayer hoppings \cite{Partoens}  between atoms of the same sublattice (A-A or B-B type), and $t_{ij}=0.12$ eV for skew interlayer hopping between atoms of different sublattices. 

\begin{figure}
\begin{tabular}{l}
\includegraphics[width=0.4\columnwidth]{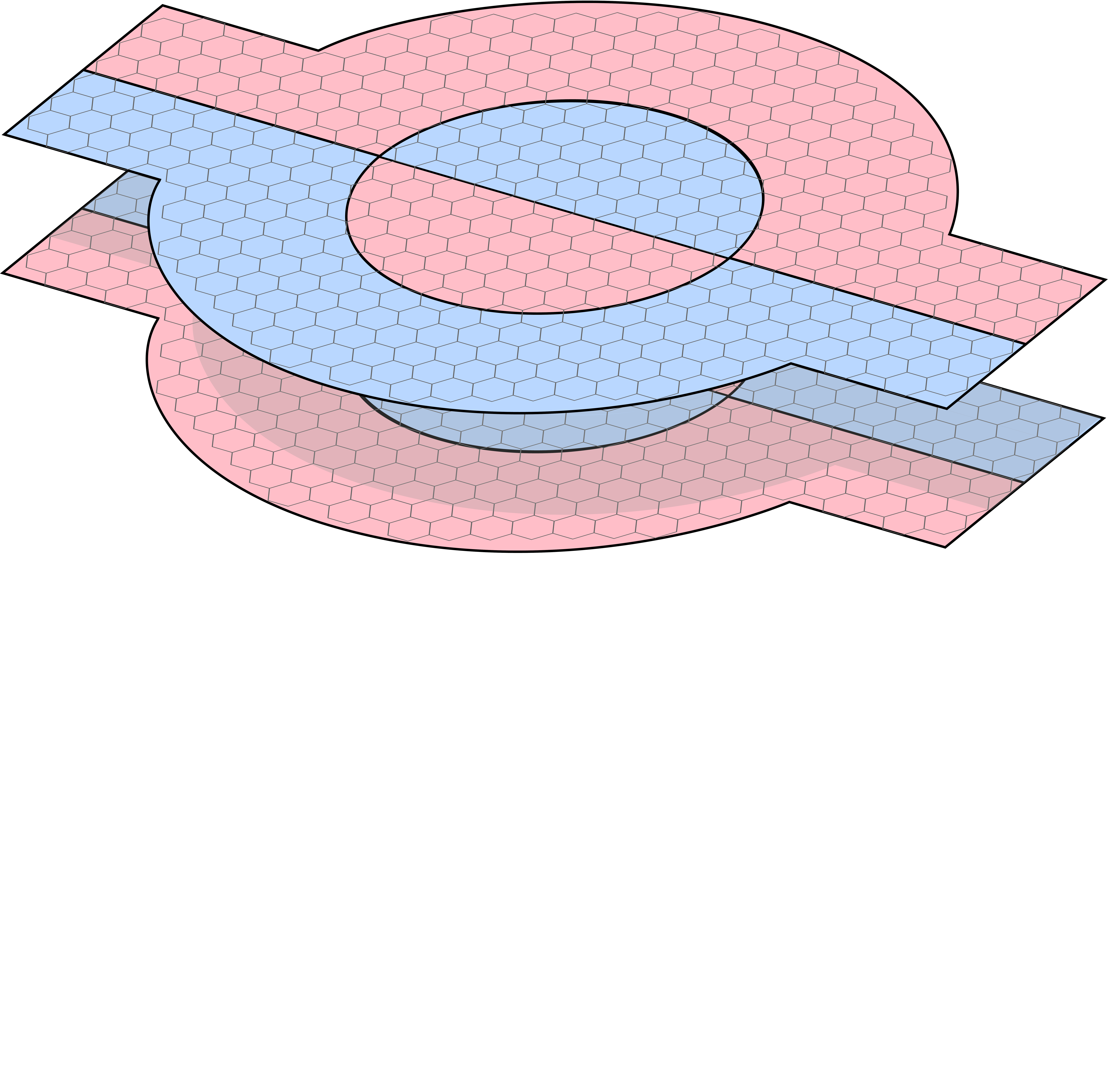}\put(1,30){(a)}
\includegraphics[width=0.6\columnwidth]{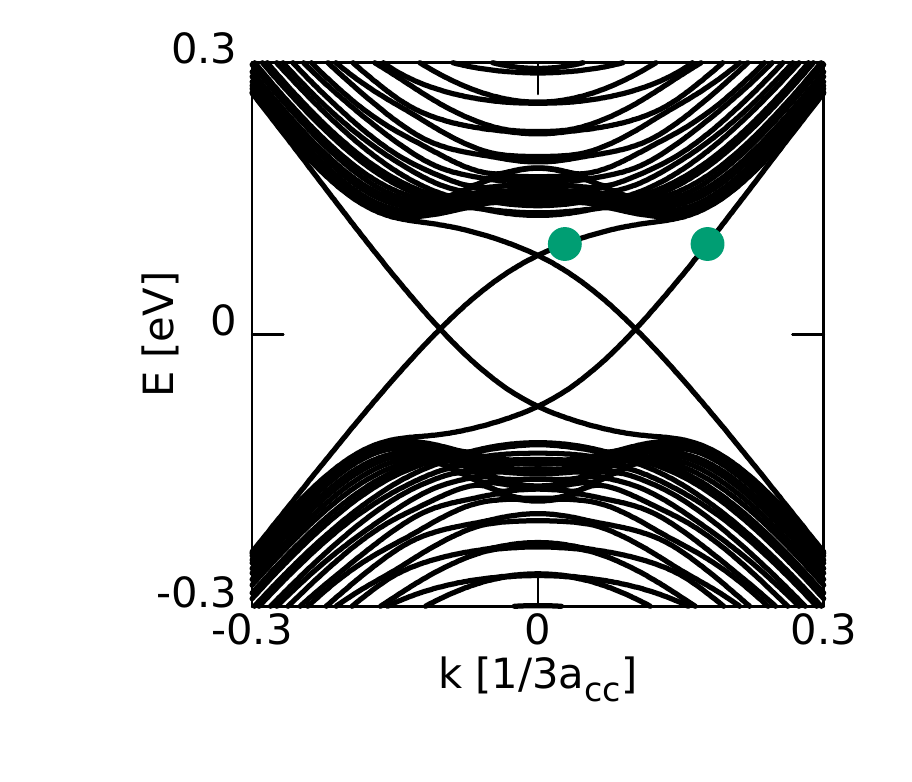} \put(1,30){(b)}\;\\
\includegraphics[width=0.8\columnwidth]{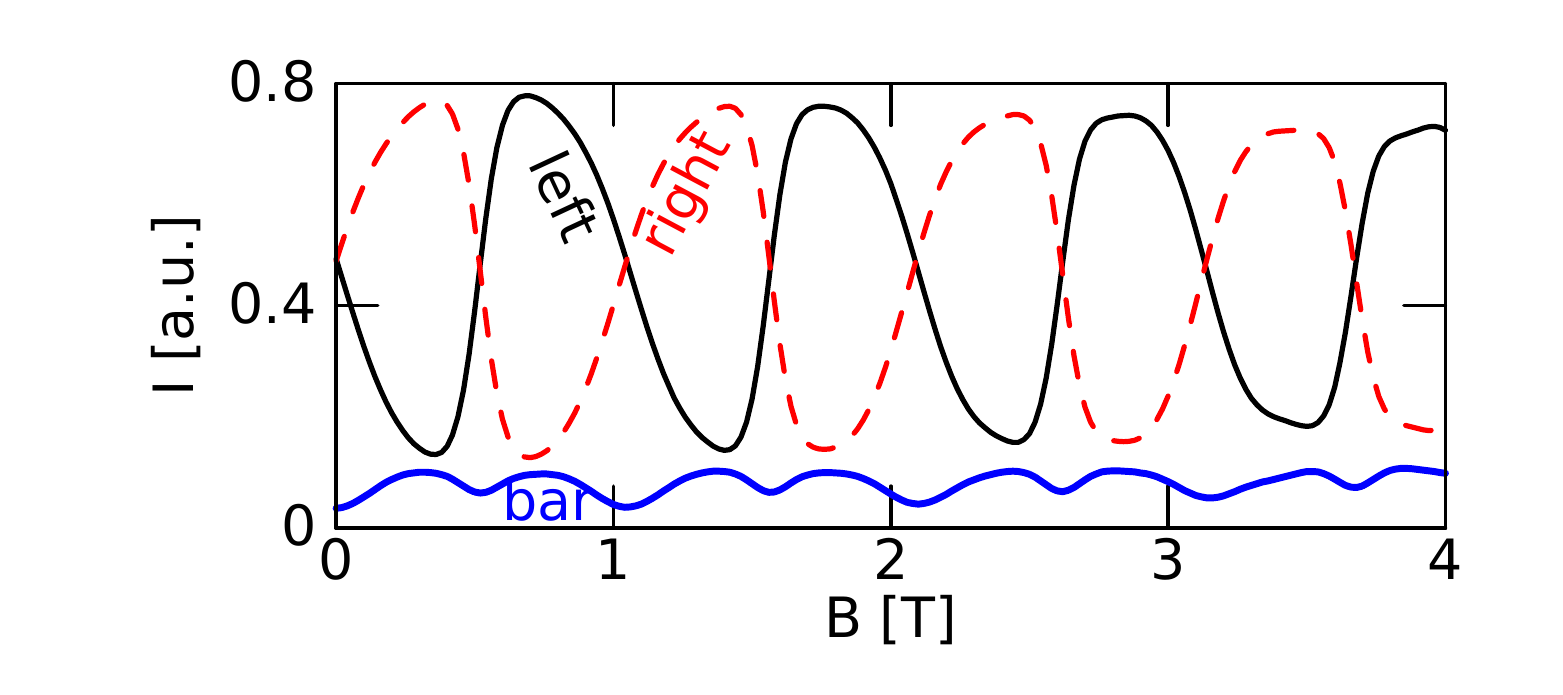} \put(1,80){(c)}\;\\
\includegraphics[width=0.8\columnwidth]{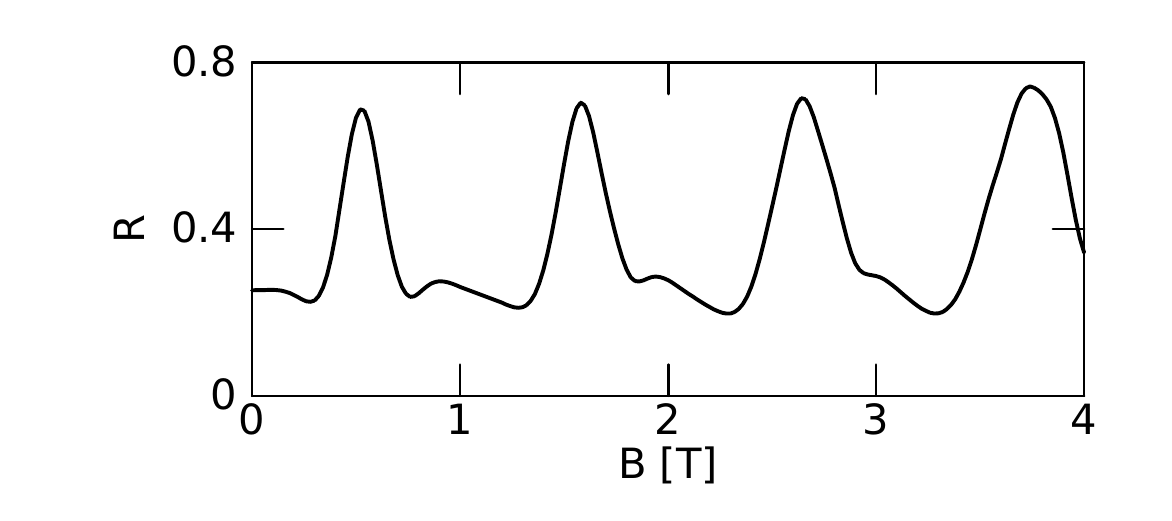}\put(1,80){(d)}\\
\end{tabular}
\caption {(a) Schematics of the potential on the upper and lower layers
of bilayer Bernal stacked graphene. Blue and red colors correspond to opposite sign of
the potential.  
We consider an armchair nanoribbon of width $W=49.2$ nm, with 1604 atoms in the elementary cell (on both layers). The zero line forms the ring of radius $R=50$ nm. The computational box in the ring area covers an area of radius  74.5 nm.
(b) The dispersion relation for the amrchair nanoribbon feeding the current to the ring. The dots indicate the Fermi wave vectors 
at the bands with positive velocity for $E_F=0.1$ eV. $a_{cc}$ is the nearest neighbor distance for graphene.
(c) Integral of the scattering density in the left and right arms of the ring and in the bar for $E_F=0.1$ eV.
(d) Backscattering probability for a vacancy: a carbon atom removed from the center of the right
arm from a position that does not form a vertical dimer with the bottom layer. 
} \label{bila}
\end{figure}

For simulation of the ring, we assume potential of the form given by Eq. (12) on the upper 
sublattice and an opposite potential on the lower sublattice [see Fig. \ref{bila}(a)].
We set $V_g=0.2$ eV, $\lambda=4$ nm as above, but  the radius of the ring is taken equal to $R=50$ nm.
The magnetic field period corresponding to the flux quantum threading the circle
of this radius is $0.53$ T.

The dispersion relation for armchair nanoribbon is displayed in Fig. \ref{bila}(b).
For calculations we take the the Fermi energy is $E_F=0.1$ eV.
For bilayer graphene we have two energy bands instead of the single one moving up the ribbon towards the ring. The integrals of the scattering density are plotted in Fig. \ref{bila}(c)
and display the periodicity with doubled period of $2\times 0.53$ T, as in silicene.

In order to produce the backscattering we removed an atom of the upper graphene layer 
from the center of the right arm. We selected an atom that does not form a vertical dimer with the lower layer.
The backscattering probability -- the sum of probabilities for each of the incident subbands --
is given in Fig. \ref{bila}(d) and display the periodicity corresponding to the flux
through half the ring, as found above for silicene.

\color{black}
\section{Summary}
We studied the dynamics of electron wave packets in buckled silicene in inhomogeneous vertical electric field
that breaks the symmetry between the sublattices using an atomistic tight-binding approach.
We have demonstrated that the line of the electric field flip in silicene supports a smooth
untrembling motion of unspreading wave packets that are topologically protected from backscattering. 
We proposed a form of a quantum ring that uses branching of the zero line to split the wave packets
and to make them interfere again. The ring stores the packet for a finite time that can be controlled
with the external magnetic field. { For short wave packets the time spent by the electron in the left and right arms of the ring is a periodic function of the flux with the period
of the flux quantum threading the ring. We found that for long packets, close to the plane waves, the electron transport across the rings becomes asymmetrical with an imbalance of the electron transfer across the left and right halves of the ring. In consequence the magnetic period is doubled.
 This  result is reproduced by stationary scattering calculations.
We demonstrated that the same effect is found for rings defined in  bilayer graphene. 
The point defects produce backscattering probability that has a period of the flux quantum threading 
ring for short packets. For long packets and in the stationary transport the period of the backscattering probability is doubled.
}

\section*{Acknowledgments}
The results of Section III.F were obtained by B.R. (stationary transport silicene) that is supported by 
by the Polish Government budget for science 2017--2021 within the Diamentowy Grant project (Grant No. 0045/DIA/2017/46).
The results of Section III.G were calculated by A.M.-K. (stationary transport bilayer graphene) that is supported by National Science Centre (NCN) according to decision DEC-2015/17/B/ST3/01161.
The rest of the results were provided by B.S. that is supported by the National Science Centre (NCN) according to decision DEC-2016/23/B/ST3/00821.
B.R. acknowledges the support of EU PhD Project POWER.03.02.00-
00-I004/16. The calculations were performed
on PL-Grid Infrastructure at ACK Cyfronet AGH.

\end{document}